\documentclass[onefignum,onetabnum]{siamonline190516}

\pdfoutput=1

\usepackage{graphics,graphicx,amsmath,amssymb,tikz,psfrag}
\usepackage{booktabs}
\usepackage{algorithmic}
\usepackage{algorithm}
\usepackage{graphicx}
\usepackage{subfigure}
\usepackage{url}
\usepackage{bm}
\usepackage{booktabs}
\usepackage{multirow}
\usepackage{rotating}
\usepackage{amssymb, amsmath}
\usepackage{xmpmulti}
\usepackage{graphicx,epstopdf}
\usepackage[english]{babel}
\usepackage{tikz}
\usepackage[customcolors]{hf-tikz}
\usetikzlibrary{patterns}
\usetikzlibrary{arrows}
\usepackage{bibentry}
\usepackage{todonotes}
\usepackage{capt-of}
\usepackage{enumitem}
\usepackage{pdfpages}

\graphicspath{ {figures/} }

\newcommand{\tp}[1]{{#1^{\hspace{0pt}\mathrm{T}}}}

\newcommand{\cov}[1]{{\operatorname{cov} ( #1 )}}
\newcommand{\var}[1]{{\operatorname{var} ( #1 )}}

\newcommand{\R}{\mathbb{R}}
\newcommand{\prob}{\mathbb{P}}
\newcommand{\N}{\mathcal{N}}

\newcommand{\y}{\bm{y}}
\newcommand{\x}{\bm{x}}

\newcommand{\w}{\bm{w}}
\newcommand{\K}{\bm{K}}
\newcommand{\0}{\bm{0}}

\newcommand{\h}{\bm{h}}
\newcommand{\I}{\bm{I}}

\newcommand{\D}{\bm{D}}

\newcommand{\A}{\bm{A}}
\newcommand{\U}{\bm{U}}
\newcommand{\V}{\bm{V}}

\newcommand{\eps}{\varepsilon}

\begin{document}

\title{Objective frequentist uncertainty quantification for atmospheric CO$_2$ retrievals\thanks{PP and MK were supported by JPL subcontract no.\ 1629749. JH's part of the research was carried out at the Jet Propulsion Laboratory, California Institute of Technology, under a contract with the National Aeronautics and Space Administration.}}

\author{Pratik Patil\thanks{Department of Statistics and Data Science and Machine Learning Department, Carnegie Mellon University, Pittsburgh, PA 15213 (\email{pratik@cmu.edu}).}
\and Mikael Kuusela\thanks{Department of Statistics and Data Science, Carnegie Mellon University, Pittsburgh, PA 15213\newline (\email{mkuusela@andrew.cmu.edu}).}
\and Jonathan Hobbs\thanks{Jet Propulsion Laboratory, California Institute of Technology, Pasadena, CA 91109\newline (\email{jonathan.m.hobbs@jpl.nasa.gov}).}}

\maketitle

\begin{abstract}
The steadily increasing amount of atmospheric carbon dioxide (CO$_2$) is affecting the global climate system and threatening the long-term sustainability of Earth's ecosystem. In order to better understand the sources and sinks of CO$_2$, NASA operates the Orbiting Carbon Observatory-2~\&~3 satellites to monitor CO$_2$ from space. These satellites make passive radiance measurements of the sunlight reflected off the Earth's surface in different spectral bands, which are then inverted in an ill-posed inverse problem to obtain estimates of the atmospheric CO$_2$ concentration. In this work, we propose a new CO$_2$ retrieval method that uses known physical constraints on the state variables and direct inversion of the target functional of interest to construct well-calibrated frequentist confidence intervals based on convex programming. We compare the method with the current operational retrieval procedure, which uses prior knowledge in the form of probability distributions to regularize the problem. We demonstrate that the proposed intervals consistently achieve the desired frequentist coverage, while the operational uncertainties are poorly calibrated in a frequentist sense both at individual locations and over a spatial region in a realistic simulation experiment. We also study the influence of specific nuisance state variables on the length of the proposed intervals and identify certain key variables that can greatly reduce the final uncertainty given additional deterministic or probabilistic constraints, and develop a principled framework to incorporate such information into our method.
\end{abstract}



\section{Introduction}
\label{sec:introduction}

Global measurements of atmospheric carbon dioxide (CO$_2$) concentration are essential for understanding Earth's carbon cycle, a key component of our planet's climate system. Space-borne observing systems provide the primary way of obtaining atmospheric CO$_2$ measurements globally at spatial and temporal resolutions useful for investigating central questions in carbon cycle science \cite{rayner2001}. A series of satellites named Orbiting Carbon Observatory-2~\&~3 (OCO-2~\&~3) \cite{ElderingAMT2017, eldering2019}, launched by NASA in July 2014 and May 2019, respectively, constitute the current state-of-the-art in space-based CO$_2$ observing systems. These instruments use the sunlight reflected off the Earth's surface to infer the CO$_2$ concentration in the atmosphere below. Since the observations are indirect measurements of the quantity of interest, the task of estimating the atmospheric state, known as \emph{retrieval} in remote sensing \cite{rodgers2000}, is an ill-posed inverse problem \cite{Kaipio2005,Engl2000,tarantola2005}. The ultimate goal of these missions is to estimate the vertically averaged atmospheric CO$_2$ concentration at high precision in order to better understand the sources and sinks of CO$_2$ in the Earth system \cite{crisp2004}.

Estimating atmospheric CO$_2$ concentrations from space is a highly nontrivial task. Designing and building the required remote sensing instrument and developing the mathematical forward model for relating the scientifically relevant quantities to the actual satellite observations are both extremely challenging tasks \cite{ODellEtAl2018}. However, statistically, the main complication arises from the fact that in order to convert the raw satellite observations into CO$_2$ concentrations, one needs to solve the associated ill-posed inverse problem \cite{cressie2018}. A satellite on low-Earth orbit is only able to measure CO$_2$ indirectly through its effect on the sunlight passing through the atmosphere. Information about CO$_2$ at different altitudes will therefore inevitably be confounded in the raw observations. Inverting the forward model to obtain a reconstruction of the atmospheric CO$_2$ profile at different altitudes will hence result in highly oscillatory and uncertain solutions which, at first glance, may seem to have little scientific value. 

The OCO-2~\&~3 Science Teams are well-aware of these challenges and the operational missions essentially employ two strategies to circumvent the forward model ill-posedness \cite{atbd2019}. First, the missions acknowledge that it is not feasible to retrieve the full vertical CO$_2$ profile from space. Instead, the missions have identified the vertically averaged CO$_2$ concentration, denoted by $X_\mathrm{CO2}$, as their primary quantity of interest, and the retrieval and validation efforts are focused on the accuracy and precision of this scalar quantity. Second, in order to estimate $X_\mathrm{CO2}$, the missions employ a strategy where first a regularized CO$_2$ profile is reconstructed (or more precisely, a regularized state vector containing the CO$_2$ profile and other retrieved atmospheric quantities), which is then used to calculate the corresponding $X_\mathrm{CO2}$ value. The regularization is achieved using a Bayesian approach where a prior distribution on the underlying state variables is used to promote physically plausible CO$_2$ profiles \cite{atbd2019, connor2016, connor2008, cressie2016}. The prior mean of the CO$_2$ profile is carefully designed to incorporate major large-scale variations in CO$_2$ over both space (latitude) and time (seasonality, long-term trends) \cite{atbd2019,ODellEtAl2018}. Even so, regional biases are found in the retrieved $X_\mathrm{CO2}$ when compared to ground-based validation sources \cite{WunchEtAl2017, kulawikval, kiel2019, wu2018}.

In this paper, we focus on rigorous uncertainty quantification for the retrieved $X_\mathrm{CO2}$. In contrast to most existing works in remote sensing, we approach the problem from the perspective of frequentist statistics. We demonstrate that the existing retrieval procedure, if evaluated using frequentist performance measures, may lead to miscalibrated uncertainties for $X_\mathrm{CO2}$ due to the intermediate regularization step. We then show that it is possible to obtain better-calibrated uncertainties by adopting an approach that avoids explicit regularization and instead directly forms an implicitly regularized confidence interval for $X_\mathrm{CO2}$. The proposed method is developed for linear or linearized forward operators, but extensions to nonlinear cases are possible.

To introduce some of the key ideas, it is worth considering a simplified version of the CO$_2$ retrieval problem. The problem is typically formulated in terms of an unknown state vector $\bm{x}$ that includes both the vertical CO$_2$ profile of interest and other geophysical nuisance variables that affect the satellite observations. Assume that the state vector $\bm{x}$ is related to the observations $\bm{y}$ by the linear model $\bm{y} = \bm{K} \bm{x} + \bm{\eps}$, where $\bm{K}$ is a known forward operator dictated by the physics of the problem and $\bm{\eps}$ represents stochastic noise in the measurement device with mean zero and covariance $\bm{\Sigma}_{\bm{\eps}}$. The fundamental challenge here is that $\bm{K}$ is an ill-conditioned matrix so that its singular values decay rapidly. Assume, for now, that $\bm{K}$ has full column rank, and therefore the least-squares estimator of $\bm{x}$ is given by $\bm{\hat{x}} = (\tp{\bm{K}} \bm{K})^{-1} \tp{\bm{K}} \bm{y}$. The covariance matrix of this estimator is $\cov{\bm{\hat{x}}} = (\tp{\bm{K}} \bm{K})^{-1} \tp{\bm{K}} \bm{\Sigma}_{\bm{\eps}} \bm{K} (\tp{\bm{K}} \bm{K})^{-1}$. Due to the ill-posedness of $\bm{K}$, the fluctuations in the noise $\bm{\eps}$ get amplified in the inversion and the estimator $\bm{\hat{x}}$ exhibits large oscillations within the CO$_2$ profile that tend to be anti-correlated from one altitude to the next. This is also reflected in the covariance $\cov{\bm{\hat{x}}}$, and any confidence intervals derived for the individual CO$_2$ elements in $\bm{x}$ based on $\cov{\bm{\hat{x}}}$ would be extremely wide, indicating, as they should, that the observations $\bm{y}$ do not contain enough information to effectively constrain CO$_2$ at a given altitude. However, this should not deter us from trying to constrain \emph{other functionals of $\bm{x}$} based on $\bm{\hat{x}}$. Of particular interest, in our case, is the vertically averaged CO$_2$ concentration given by the functional $X_\mathrm{CO2} = \tp{\bm{h}}\bm{x}$, where $\bm{h}$ is a known vector of weights. The plug-in estimator of $X_\mathrm{CO2}$ is $\hat{X}_\mathrm{CO2} = \tp{\bm{h}}\bm{\hat{x}}$ with variance $\var{\hat{X}_\mathrm{CO2}} = \tp{\bm{h}} \cov{\bm{\hat{x}}} \bm{h}$. Since the mapping from $\bm{x}$ to $X_\mathrm{CO2}$ is an averaging operation, one would expect that the anti-correlated fluctuations in the unregularized $\bm{\hat{x}}$ largely cancel out as it is mapped into $\hat{X}_\mathrm{CO2}$, resulting in a well-behaved estimator of $X_\mathrm{CO2}$, as also suggested by the results in \cite{ramanathan2018}. When the noise $\bm{\eps}$ is Gaussian, which is a good approximation here, one can then use the variance $\var{\hat{X}_\mathrm{CO2}}$ to construct a frequentist confidence interval around~$\hat{X}_\mathrm{CO2}$. Assuming that the forward model is correctly specified, these intervals have guaranteed frequentist coverage for $X_\mathrm{CO2}$, \emph{without requiring any additional information about~$\bm{x}$} (e.g., information about smoothness or specification of a prior distribution). Arguably, these intervals provide an objective measure of uncertainty of $X_\mathrm{CO2}$ in the absence of specific prior information about $\bm{x}$.

The actual retrieval problem is more complex than the simplified situation described above. First, the forward operator relating the state vector $\bm{x}$ to the observations $\bm{y}$ is a nonlinear function of $\bm{x}$ \cite{atbd2019}. Second, there are known physical constraints on the state vector $\bm{x}$ that should ideally be taken into account in the retrieval. For example, those elements of $\bm{x}$ that correspond to CO$_2$ concentrations should be constrained to be non-negative. Third, the forward mapping need not be injective. This means, for example, that the matrices corresponding to a linearization of the forward mapping may be rank deficient. In this paper, we address these last two complications in the case of a linearized approximation to the nonlinear forward operator. In other words, we seek to rigorously quantify the uncertainty of $X_\mathrm{CO2} = \tp{\bm{h}}\bm{x}$ under the model $\bm{y} = \bm{K} \bm{x} + \bm{\eps}$, where $\bm{K}$ need not have full column rank, $\bm{x} \in C$, where $C$ is a set of known physical constraints (i.e., constraints that hold with probability 1), and $\bm{\varepsilon}$ is noise with a known Gaussian distribution. We focus on the case of affine constraints for the elements of the state vector~$\bm{x}$, and in particular, on non-negativity constraints for certain elements of the state vector. Under this setup, we seek to construct $(1-\alpha)$ frequentist confidence intervals for $X_\mathrm{CO2}$ without imposing any other regularization on $\bm{x}$. We propose a procedure that is demonstrated to consistently provide nearly nominal $(1-\alpha)$ frequentist coverage, including in situations where the existing retrieval procedure can be severely miscalibrated. Even though our procedure relies on much weaker assumptions, the new intervals are not excessively wide as the problem is implicitly regularized by the choice of the functional $\tp{\bm{h}}\bm{x}$ and the constraints $\bm{x} \in C$.

We also investigate the potential implications of these results on CO$_2$ flux estimates \cite{eldering2017} by studying the behavior of the different methods over a small spatial domain. We find that in the existing operational retrievals, the interaction between the regularizing prior and the spatially dependent true state vectors can lead, at least in the specific example studied, to a situation where the miscalibration of the $X_\mathrm{CO2}$ intervals varies in a spatially coherent fashion. As a result, the reported uncertainties can be systematically too small or too large over a given spatial region. It is possible that retrievals with such uncertainties could lead to spurious CO$_2$ flux estimates in downstream analyses. On the other hand, the sampling properties of our proposed intervals do not vary spatially, which makes them potentially more suitable for downstream scientific use.

In addition, we study the contributions of individual state vector elements to the $X_\mathrm{CO2}$ uncertainty, identifying surface pressure and a certain aerosol variable as the key parameters that contribute most to the final uncertainty. This means that the $X_\mathrm{CO2}$ uncertainty could potentially be further reduced if additional external information was available to constrain these two variables. We provide a principled framework for incorporating such information in either deterministic or probabilistic forms into our method and investigate the extent to which such additional information on surface pressure reduces the $X_\mathrm{CO2}$ uncertainty.

This work relates to a wider discussion on uncertainty quantification in ill-posed inverse problems (see, e.g., \cite{stark2015, stark2011, tenorio2017, stuart2010}). In applied situations, uncertainty quantification in inverse problems tends to be dominated by Bayesian approaches that regularize the problem using a prior distribution. This is certainly the case in atmospheric sounding \cite{rodgers2000}, but also in other domain sciences (e.g., \cite{Biegler2011,Kaipio2005,Isaac2015,Martin2012,vandyk2006,Weir1997}). Penalized frequentist techniques, such as penalized maximum likelihood or Tikhonov regularization (also known as ridge regression~\cite{hoerl1970}), are closely related to Bayesian approaches since one can usually interpret the penalty term as a Bayesian log-prior (see, e.g., Sections~7.5 and 7.6 in \cite{murphy2012}). These techniques, in which the problem is explicitly regularized, are challenging from the perspective of frequentist uncertainty quantification since intervals centered around a regularized point estimator tend to be systematically offset from the true value of the unknown quantity due to the bias in the point estimator that regularizes the problem. This bias has been investigated in multiple remote sensing retrieval settings \cite{rodgers2000, nguyen2019, MaahnBAMS} and has been shown to lead to drastic undercoverage for the intervals in other applied situations \cite{kuusela2015, kuusela2017, kuusela2016}. There exists, however, a lesser-known line of work (see \cite{stark1992, rust1994, rust1972} and the references therein) that attempts to construct truly frequentist confidence intervals in ill-posed problems without relying on explicitly regularized point estimators. One of the key ideas is to use physically known objective constraints to regularize the problem instead of a subjective prior distribution or a penalty term. This enables deriving intervals with \emph{guaranteed frequentist coverage}  \cite{stark1992, stark1995}. This paper builds upon these ideas, but adds to the discussion by highlighting the important role of the functional of interest in implicitly regularizing the problem. We also focus on intervals which are designed to constrain one functional at a time, in contrast to some previous techniques \cite{stark1992, kuusela2017} that provide simultaneous intervals for \emph{all} functionals at once, which leads to substantial overcoverage if only one or a small subset of functionals is needed.

The rest of this paper is organized as follows. To set up the problem, we briefly describe the physics of observing CO$_2$ from space and the corresponding statistical model in Section~\ref{sec:setup}. We then describe the proposed frequentist uncertainty quantification method and discuss its properties in Section~\ref{sec:proposed}. Next, we outline the existing operational retrieval procedure and analyze its properties in Section~\ref{sec:operational}.  Section~\ref{sec:numerical} compares the coverage performance of the operational and proposed procedures both for an individual sounding location and over a small spatial region using simulated data from a realistic generative model. In Section~\ref{sec:extension}, we further investigate the proposed method to better understand the impact of the individual state vector elements on the final interval length, and provide a framework to incorporate additional deterministic or probabilistic constraints into our method. Finally, Section~\ref{sec:conclusion} offers concluding remarks and directions for future work. The Appendices and the Supplement \cite{Patil2021Supp} contain derivations and other supplementary results.

\section{Problem background and setup}
\label{sec:setup}

\subsection{Remote sensing of carbon dioxide} \label{sec:co2sensing}

\begin{figure}
	\label{fig:retrieval_setup}
	\centering
	\includegraphics[width=0.6\columnwidth]{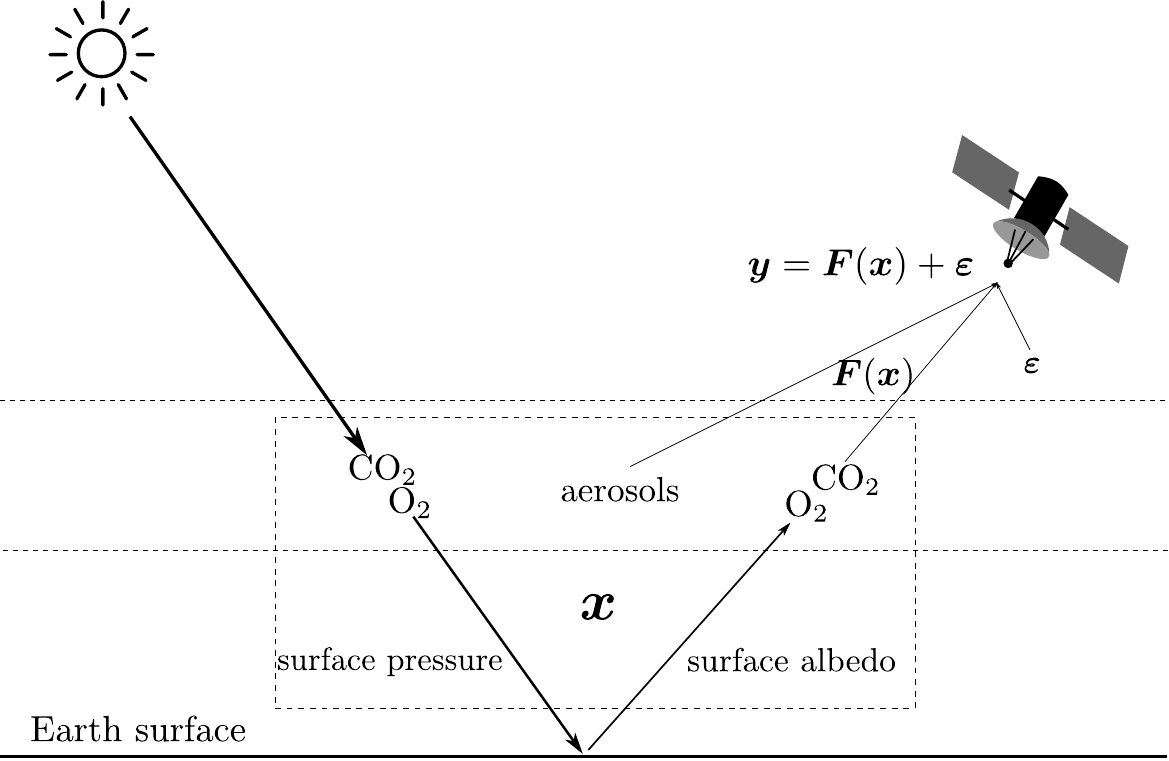}
	\caption{Schematic of space-based CO$_2$ sensing in the OCO-2 mission.}
\end{figure}

Remote sensing of atmospheric CO$_2$ is feasible due to the absorption of solar radiation by CO$_2$ molecules at specific wavelengths, particularly in the infrared (IR) portion of the electromagnetic spectrum. In this part of the spectrum, variations in the observed top-of-the-atmosphere radiation can also be induced by other surface and atmospheric properties, including albedo (surface reflectivity), absorption by other atmospheric trace gases, and absorption and scattering in the presence of clouds and aerosol particles. These processes are illustrated schematically in Figure~\ref{fig:retrieval_setup}. These additional effects explain most of the variation in the radiance (intensity of the observed radiation) that is seen by a downward looking satellite at the top of the atmosphere. Radiance changes due to variation in CO$_2$ are more subtle. CO$_2$-focused remote sensing instruments, such as OCO-2~\&~3, therefore require high-precision radiance observations at fine spectral resolution. The OCO-2 and OCO-3 instruments are duplicates of the same design. Each instrument includes three imaging grating spectrometers that each correspond to a narrow IR band. These are the O$_2$ A-band centered around 0.765~$\mu$m, the weak CO$_2$ band centered around 1.61~$\mu$m, and the strong CO$_2$ band centered near 2.06~$\mu$m. The O$_2$ A-band includes numerous absorption lines for atmospheric O$_2$, and the two CO$_2$ bands include absorption lines for CO$_2$~\cite{atbd2019}.

A collection of observed radiances at a particular time and location is known as a \emph{sounding}. For OCO-2 \& 3, a sounding includes 1016 radiances in each of the three spectral bands. Figure~S1 in the Supplement \cite{Patil2021Supp} depicts an example sounding for OCO-2. The fine wavelength spacing within each band ensures the ability to resolve individual absorption features. Since atmospheric O$_2$ has a nearly constant fractional abundance of 0.209, the absorption in the O$_2$ A-band can be used to estimate the total amount of dry air in the atmospheric column, which is sometimes termed as the radiative path length. In the retrieval, this is formally represented by retrieving the atmospheric surface pressure. This can be combined with the absolute absorption in the CO$_2$ bands to estimate the relative abundance, or dry-air mole fraction, of CO$_2$. In addition, the A-band in particular has sensitivity to cloud and aerosol scattering, which are also estimated in the retrieval process.

While the instruments themselves are nearly identical, OCO-2 and OCO-3 have different observing patterns. OCO-2 is in a polar orbit as part of a satellite constellation known as the A-train with observations collected exclusively in the early afternoon local time \cite{atbd2019}. OCO-3 has recently been installed on the International Space Station (ISS) and is collecting observations in tropical and mid-latitude regions at varying times of the day following the ISS precessing orbit~\cite{eldering2019}. In the rest of this paper, we primarily focus on OCO-2, but we expect our conclusions to also apply to OCO-3 due to the similarity of the two instruments.

\subsection{Mathematical model}
\label{sec:mathematical_model}

The physical model for CO$_2$ remote sensing as illustrated in Figure~\ref{fig:retrieval_setup} can be mathematically written down as
\begin{equation}
	\bm{y} = \bm{F}(\bm{x}) + \bm{\eps}, \label{eq:fullForwardModel}
\end{equation}
where $\bm{x} \in \mathbb{R}^p$ is an unknown state vector, $\bm{y} \in \mathbb{R}^n$ is a vector of observed radiances,\linebreak $\bm{F}: \mathbb{R}^p \rightarrow \mathbb{R}^n$ models the physical processes described in Section~\ref{sec:co2sensing} that relate the state vector to the expected radiances and $\bm{\eps}$ represents zero-mean instrument noise. The noise is assumed to have a Gaussian distribution, $\bm{\eps} \sim \mathcal{N}(\mathbf{0}, \bm{\Sigma}_{\bm{\eps}})$, with a known diagonal covariance matrix $\bm{\Sigma}_{\bm{\eps}}$. For the OCO\nobreakdash-2 mission, we have $n \gg p$. Despite this, the problem of retrieving $\bm{x}$ based on $\bm{y}$ is badly ill-posed due to poor conditioning of $\bm{F}$.

The ultimate goal of the retrieval is to estimate a certain functional of the state vector $\theta(\bm{x}) \in \mathbb{R}$ using the observations $\bm{y}$. We assume that the functional of interest is linear so it can be written in the form $\theta = \bm{h}^T \bm{x}$, where the weights $\bm{h}$ are assumed to be known. We specifically focus on $\theta$ corresponding to $X_\mathrm{CO2}$, the column-averaged CO$_2$ concentration at the sounding location, which, to a good approximation, is of this form.

The state vector $\bm{x}$ contains all the physical quantities that are thought to affect the radiance measurement $\bm{y}$. This includes the vertical CO$_2$ concentrations, but also other geophysical quantities, as outlined in Section~\ref{sec:co2sensing} and described in more detail in Section~\ref{sec:numerical_model}. Statistically, these other quantities can be understood as nuisance variables since the functional of interest does not directly depend on them ($h_i = 0$ for these variables).

The actual full-physics forward operator $\bm{F}$ is a non-linear map from $\mathbb{R}^p$ to $\mathbb{R}^n$ \cite{atbd2019}. This complication can detract from the fundamental challenges involved in quantifying the uncertainty of the retrievals. In order to be able to focus on the key statistical issues, we linearize in this work the forward operator $\bm{F}(\bm{x})$ at a particular $\bm{x} = \bm{x}'$ such that $\bm{F}(\bm{x}) \approx \bm{K} \bm{x} + \bm{F}(\bm{x}') - \bm{K} \bm{x}'$, where $\bm{K}  = \frac{\partial \bm{F}(\bm{x})}{\partial \bm{x}}\big|_{\bm{x} = \bm{x}'}$ is the Jacobian of $\bm{F}$ evaluated at~$\bm{x}'$. This differs from the OCO-2 operational retrieval method which takes the nonlinearity of the forward operator into account. Even so, the operational uncertainty estimate uses a linearization about the final solution \cite{atbd2019}.

Putting these elements together, we have the following Gaussian linear model:
\begin{equation}
\label{eq:linear_forward_model}
	\bm{y}' = \bm{K} \bm{x} + \bm{\eps}, \qquad \bm{\eps} \sim \mathcal{N}(\mathbf{0}, \bm{\Sigma}_{\bm{\eps}}),
\end{equation}
where $\bm{y}' = \bm{y} - \bm{F}(\bm{x}') + \bm{K} \bm{x}'$. To simplify the notation, we denote $\bm{y}'$ as $\bm{y}$ in the rest of this paper. Under this model, our goal is to obtain a $(1-\alpha)$ confidence interval of the form $\left[\underline{\theta}, \overline{\theta}\right]$ for the functional $\theta = \bm{h}^T \bm{x}$  with the frequentist coverage guarantee $\prob_{\bm{\varepsilon}}(\theta \in \left[\underline{\theta}, \overline{\theta}\right]) \approx 1 - \alpha$ for any~$\bm{x}$, where $(1 - \alpha)$ is the desired confidence level and the probability statement is with respect to the distribution of the noise $\bm{\varepsilon}$.

\section{Proposed frequentist retrieval procedure}
\label{sec:proposed}

\subsection{Motivation}

The key idea of our proposed method is to let known physical constraints and the functional of interest regularize the problem without imposing other external a priori beliefs about the state elements. We demonstrate using simulations that this suffices for obtaining well-calibrated and reasonably sized confidence intervals, as long as the constraints hold with probability 1 and the functional is an operation, such as averaging or smoothing, that tends to reduce noise. The procedure is formulated in terms of convex optimization problems that find the upper and lower endpoints of the confidence interval \cite{rust1972,rust1994,stark1992}. Below, we first describe the procedure, followed by a brief analysis of its properties. We provide two complementary perspectives on the method, one from the point of view of optimization in the state space $\R^p$ and another from the dual perspective of optimization in the radiance space~$\R^n$.

\subsection{Method outline} \label{sec:proposalOutline}

Unlike the operational procedure described in detail in Section~\ref{sec:operational}, our proposed method directly constructs confidence intervals for the functional of interest $\theta = \bm{h}^T \bm{x}$. More specifically, the goal is to construct a $(1-\alpha)$ confidence interval $[\underline{\theta},\overline{\theta}]$ for $\theta$ under the model $\bm{y} \sim \mathcal{N}(\bm{K}\bm{x},\bm{I})$ subject to external information on $\bm{x}$ in the form of the affine constraint $\bm{A}\bm{x} \leq \bm{b}$ and without requiring $\bm{K}$ to have full column rank. The linear forward model in Equation~\eqref{eq:linear_forward_model} can always be transformed into this form by taking the Cholesky factorization $\bm{\Sigma}_{\bm{\varepsilon}} = \bm{L}\bm{L}^T$, calculating $\bm{\tilde{y}} = \bm{L}^{-1} \bm{y}$ and $\bm{\tilde{K}} = \bm{L}^{-1}\bm{K}$, and then redefining $\bm{y} \leftarrow \bm{\tilde{y}}$ and $\bm{K} \leftarrow \bm{\tilde{K}}$. We assume throughout the remainder of this section that this transformation has been applied to the model. The matrix $\A$ and the vector $\bm{b}$ can encode various types of affine constraints on the state vector elements: for example, non-negativity constraints for individual elements of $\bm{x}$, two-sided bounds for individual elements of $\bm{x}$, or affine constraints involving multiple elements of $\bm{x}$ at once. The endpoints of the interval $[\underline{\theta},\overline{\theta}]$ are obtained as the objective function values of two convex optimization problems. The convex programs are chosen so that the coverage $\prob_{\bm{\varepsilon}}(\theta \in \left[\underline{\theta}, \overline{\theta}\right])$ is as close as possible to the nominal value $(1-\alpha)$ for all $\bm{x}$ satisfying the constraint $\bm{A}\bm{x} \leq \bm{b}$.

\subsubsection{Primal point of view} \label{sec:proposalPrimal}

The lower endpoint $\underline{\theta}$ is the optimal objective function value of the following minimization problem \cite{rust1972, rust1994}:
\begin{equation}\label{eq:primalMin}
\begin{array}{ll}
\mbox{minimize}   & \h^T \x \\
\mbox{subject to} & \| \y  - \K \x \|^2 \le z_{1- \alpha/2}^2 + s^2, \\
& \A \x \le \bm{b},
\end{array}
\end{equation}
where $z_{1 - \alpha/2}$ is the $(1 - \alpha/2)$ standard normal quantile and the slack factor $s^2$ is defined as the objective function value of the following program:
\begin{equation}\label{eq:program_for_s}
\begin{array}{ll}
\mbox{minimize}   & \| \y  - \K \x \|^2 \\
\mbox{subject to} & \A \x \le \bm{b}.
\end{array}
\end{equation}
The upper endpoint $\overline{\theta}$ is the optimal value of a similar maximization problem:
\begin{equation}\label{eq:primalMax}
\begin{array}{ll}
\mbox{maximize}   & \h^T \x \\
\mbox{subject to} & \| \y  - \K \x \|^2 \le z_{1- \alpha/2}^2 + s^2, \\
& \A \x \le \bm{b},
\end{array}
\end{equation}
where $s^2$ is again given by program \eqref{eq:program_for_s}.

To explain the intuition behind this construction, we start with the approach described in \cite{stark1992}. Consider the two sets $D = \{ \bm{x} \in \R^p : \| \y - \K \x \|^2 \le \chi^2_{n,1- \alpha} \}$, where $\chi^2_{n,1- \alpha}$ is the $(1-\alpha)$ quantile of the $\chi^2$ distribution with $n$ degrees of freedom, and $C = \{ \bm{x} \in \R^p : \linebreak \A \x \le \nobreak \bm{b} \}$. Here $D$ is a $(1-\alpha)$ confidence set for the entire state vector $\x$ and the set $C$ encodes the feasible set of $\x$ given the constraints. Therefore, the set $C \cap D$ is also a $(1-\alpha)$ confidence set for $\x$. We can then use this confidence set to obtain a $(1-\alpha)$ confidence interval for the functional $\theta = \h^T \x$ by simply finding the extremal values of the functional over $C \cap D$ \cite{stark1992}, which corresponds to problems \eqref{eq:primalMin} and \eqref{eq:primalMax} with $z_{1- \alpha/2}^2 + s^2$ replaced by $\chi^2_{n,1- \alpha}$. However, since this choice of $D$ guarantees coverage for the entire vector $\x$, this construction produces simultaneously valid confidence intervals for any arbitrarily large collection of functionals of~$\x$. Thus, for the one particular functional we primarily care about, it produces valid but typically excessively wide intervals that are likely to have substantial overcoverage. The idea of the method above therefore is to shrink the set $D$ by calibrating the radius appropriately. It is suggested in \cite{rust1972, rust1994} that the appropriate radius for one-at-a-time coverage, i.e., for obtaining coverage for a single target functional, is $z_{1- \alpha/2}^2 + s^2$, where $s^2$ is the objective function value of program \eqref{eq:program_for_s}. We will use this radius throughout the rest of this paper. One of our goals will be to study the validity of this choice and in particular to illustrate that the intervals defined by \eqref{eq:primalMin} and \eqref{eq:primalMax} are indeed well-calibrated in the $X_\mathrm{CO2}$ retrieval~problem.

When we calculate the intervals in practice, we improve the computing time by using a simplification of \eqref{eq:primalMin}--\eqref{eq:primalMax} that allows us to replace these programs by equivalent optimization problems involving $p$-variate norms instead of $n$-variate norms; see Appendix~\ref{app:comp_simplification} for details. These simplified problems are then solved using the interior-point solvers in Matlab~2019a.

\subsubsection{Dual point of view}

To gain more insight into this construction, we next look at the Lagrangian dual \cite{boyd2004} of problems \eqref{eq:primalMin} and \eqref{eq:primalMax}. When the optimal objective function value of the dual program equals that of the primal program, the problem is said to satisfy strong duality. Since programs \eqref{eq:primalMin} and \eqref{eq:primalMax} are convex, strong duality is guaranteed if the norm constraint in \eqref{eq:primalMin} and \eqref{eq:primalMax} is strictly feasible (Eq.\ (5.27) in \cite{boyd2004}). This is true if we assume that the minimizer of the slack problem \eqref{eq:program_for_s} is attained, as any such minimizer is strictly feasible for the norm constraint.

Then, the lower endpoint $\underline{\theta}$ can also be obtained as the objective function value of the following program, which is derived starting from \eqref{eq:primalMin} in Appendix~\ref{app:dual_derivation}:
\begin{equation}
\begin{array}{ll}
\mbox{maximize}   & \w^T \y - \sqrt{z_{1- \alpha/2}^2 + s^2} \| \w \| - \bm{b}^T \bm{c}\\
\mbox{subject to} & \h + \A^T \bm{c} - \K^T \w = \bm{0}, \\
& \bm{c} \ge \bm{0},
\end{array} \label{eq:dualLb}
\end{equation}
where the optimization is over the variables $\w \in \R^n$ and $\bm{c} \in \R^q$, with $q$ the number of affine constraints on $\bm{x}$, and $s^2$ is as defined above. The upper endpoint $\overline{\theta}$ is given by an analogous program which is dual to \eqref{eq:primalMax}:
\begin{equation}
\begin{array}{ll}
\mbox{minimize}   & \w^T \y + \sqrt{z_{1 - \alpha/2}^2 + s^2} \| \w \| + \bm{b}^T \bm{c} \\
\mbox{subject to} & \h - \A^T \bm{c} - \K^T \w = \bm{0}, \\
& \bm{c} \ge \bm{0}.
\end{array} \label{eq:dualUb}
\end{equation}
The dual perspective provides us more insight into the proposed interval $\left[\underline{\theta}, \overline{\theta}\right]$. To see this, consider the interval
\begin{equation}
\left[ \underline{\w}^T \y - z_{1 - \alpha/2} \| \underline{\w} \| - \bm{b}^T \underline{\bm{c}},\, \overline{\w}^T \y + z_{1 - \alpha/2} \| \overline{\w} \| + \bm{b}^T \overline{\bm{c}} \right].
\end{equation}
As we show below and in Appendix \ref{sec:dualCoverage}, if $(\underline{\w},\underline{\bm{c}})$ and $(\overline{\w},\overline{\bm{c}})$ are any fixed elements of $\mathbb{R}^n \times \mathbb{R}^q$ satisfying the constraints in programs \eqref{eq:dualLb} and \eqref{eq:dualUb}, respectively, then the above interval has correct coverage $(1-\alpha)$. This is true even when $\K$ is rank deficient and under the constraint $\A \x \le \bm{b}$ for~$\x$. Therefore, it makes sense to find $(\underline{\w},\underline{\bm{c}})$ and $(\overline{\w},\overline{\bm{c}})$ within the appropriate constraint sets such that the lower endpoint is maximized and the upper endpoint is minimized so that the overall interval is as short as possible. This optimized interval would have correct coverage if the optimized variables did not depend on $\y$, but unfortunately that is not the case here. In order to account for this optimism, it is necessary to inflate the interval to preserve coverage. The method proposed in \cite{rust1972, rust1994}, and further considered here, does this by replacing $z_{1 - \alpha/2}$ with $\sqrt{z_{1- \alpha/2}^2 + s^2}$, where $s^2$ is the slack defined above.

\subsection{Method properties} We can show the following properties for the proposed method:

\begin{itemize}
	\item \emph{Coverage}:	
	The dual formulation enables us to gain some understanding of the coverage of the proposed interval. Consider a lower endpoint of the form $\underline{\theta} = \w^T \y - z_{1- \alpha/2} \| \w \| - \bm{b}^T \bm{c}$ for some fixed $\w$ and $\bm{c}$ satisfying the constraints in program \eqref{eq:dualLb}. As shown in Appendix~\ref{sec:dualCoverage}, we can bound the miscoverage probability to obtain $\prob_{\bm{\eps}}(\underline{\theta} \ge \theta) \le \alpha/2$. Similarly, for an upper endpoint of the form $\overline{\theta} = \w^T \y + z_{1- \alpha/2} \| \w \| + \bm{b}^T \bm{c}$, where $\w$ and $\bm{c}$ are fixed and satisfy the constraints in program~\eqref{eq:dualUb}, we have $\prob_{\bm{\eps}}(\theta \ge \overline{\theta}) \le \alpha/2$. Combining the two, we have $\prob_{\bm{\eps}}(\underline{\theta} \le \theta \le \overline{\theta}) \ge 1 - \alpha$, giving the desired coverage probability. Notice, however, that when we optimize over $\w$ and~$\bm{c}$, the optimized variables will depend on the observations~$\y$ and the proof in Appendix~\ref{sec:dualCoverage} no longer holds. To account for this, the method introduces the slack factor $s^2$ to inflate the interval. Proving that the inflated interval has correct coverage is nontrivial since the slack $s^2$ itself is also a function of $\y$, but we demonstrate empirically in Section~\ref{sec:numerical} that the coverage is consistently very close to the desired value $(1-\alpha)$.
	
	\item \emph{Length}:
	Since the optimization problems defining the interval depend on the observed data $\y$, these intervals can have variable length. Our experiments in Section~\ref{sec:numerical} confirm that the interval lengths indeed do vary across $\y$ realizations, but, in our experimental setup at least, the average length does not appear to change much across different $\x$.
	
	\item \emph{Connection with classical intervals:} In the special case where $\bm{K}$ has full column rank, i.e., $\mathrm{rank}(\bm{K}) = p$, and there are no constraints on $\bm{x}$, the proposed interval reduces to the usual Gaussian standard error interval induced by the unregularized least-squares estimator of $\x$. That is, in this special case, the solutions of problems \eqref{eq:primalMin} and \eqref{eq:primalMax} yield the interval $[\hat{\theta}_\mathrm{LS} - z_{1 - \alpha/2} \,\mathrm{se}(\hat{\theta}_\mathrm{LS}), \hat{\theta}_\mathrm{LS} + z_{1 - \alpha/2} \,\mathrm{se}(\hat{\theta}_\mathrm{LS})]$, where $\hat{\theta}_\mathrm{LS} = \h^T \hat{\x}_\mathrm{LS}$ is the induced estimator of $\theta$, $\hat{\x}_\mathrm{LS} = (\K^T \K)^{-1} \K^T \y$ is the unregularized least-squares estimator of $\x$ and $\mathrm{se}(\hat{\theta}_\mathrm{LS}) = \sqrt{\h^T (\K^T\K)^{-1} \h}$ is the standard error of~$\hat{\theta}_\mathrm{LS}$. The proof is given in Appendix~\ref{app:reduction}. By standard arguments, this interval has exact $(1-\alpha)$ coverage and will have reasonable length when the mapping $\x \mapsto \theta$ acts as an implicit regularizer. In this special case, the interval has fixed length. When $\K$ is rank deficient and/or there are constraints on $\x$, the classical interval no longer applies, but the proposed interval does. The proposed interval can therefore be seen as an extension of the classical unregularized interval to these more complex~situations.
\end{itemize}

\subsection{Commentary}

The proposed method takes advantage of the fact that certain functionals themselves provide enough regularity so that we can retrieve them with reasonably sized confidence intervals given only objectively known physical constraints and without having to use additional subjective knowledge. This way the method avoids dependence on subjective external beliefs for the coverage guarantees. In practice, these intervals tend to be better calibrated but longer than the operational intervals which rely on such subjective knowledge. The interval length can be improved if additional objective information about the state variables is available to shrink the constraint set. This information could come either in the form of additional hard constraints or in the form of soft constraints of coverage statements for some of the unknown variables. Since this method is designed to satisfy a frequentist coverage statement, it is possible to combine these different uncertainties to obtain a valid, shorter interval in the end. These extensions are explored in Section \ref{sec:extension}.

\section{Existing operational retrieval procedure}
\label{sec:operational}

\subsection{Motivation}

  The existing OCO-2 operational retrieval procedure is based on a \linebreak Bayesian maximum a posteriori estimator \cite{atbd2019, rodgers2000}, where the key idea is to let a prior distribution on the state vector $\bm{x}$ regularize the problem. In remote sensing literature, this approach is called ``optimal estimation'' \cite{rodgers2000}, although optimality here depends on the choice of a cost function and typically assumes that the prior is correctly specified. We describe below the operational retrieval for our simplified setup with a linearized forward model and analyze its frequentist properties. In the actual Full Physics operational retrievals with a nonlinear forward operator, finding the maximum of the posterior is a nonlinear optimization problem which is solved using the iterative Levenberg--Marquardt~algorithm \cite{atbd2019}.
  
\subsection{Method outline}

The existing operational method for estimation and uncertainty quantification assumes a Gaussian prior distribution on the state vector, $\bm{x} \sim \mathcal{N}(\bm{\mu}_a, \bm{\Sigma}_a)$, where $\bm{\mu}_a$ and $\bm{\Sigma}_a$ are the prior mean and covariance, respectively.
The posterior under this assumption and the linear forward model \eqref{eq:linear_forward_model} is also Gaussian and is given by
\begin{equation}
	\bm{x} | \bm{y} \sim \mathcal{N}\big((\bm{K}^T \bm{\Sigma}_\eps^{-1} \bm{K} + \bm{\Sigma}_a^{-1})^{-1} (\bm{K}^T \bm{\Sigma}_\eps^{-1} \bm{y} + \bm{\Sigma}_a^{-1} \bm{\mu}_a), (\bm{K}^T \bm{\Sigma}_\eps^{-1} \bm{K} + \bm{\Sigma}_a^{-1})^{-1}\big).
\end{equation}
The point estimator $\hat{\bm{x}}$ of $\bm{x}$ is chosen to be the maximizer of the posterior distribution,
$$
\hat{\bm{x}} = (\bm{K}^T \bm{\Sigma}_\eps^{-1} \bm{K} + \bm{\Sigma}_a^{-1})^{-1} (\bm{K}^T \bm{\Sigma}_\eps^{-1} \bm{y} + \bm{\Sigma}_a^{-1} \bm{\mu}_a),
$$
which in our simplified setup is also the posterior mean. Recalling that $\bm{y} = \bm{K}\bm{x} + \bm{\varepsilon}$, this estimator can be written as a sum of three terms, $\hat{\bm{x}} = \bm{A} \bm{x} + \big ( \bm{I} - \bm{A} \big) \bm{\mu}_a + \bm{G} \bm{\eps}$,
where $\bm{G} = \nobreak (\bm{K}^T \bm{\Sigma}_\eps^{-1} \bm{K} + \bm{\Sigma}_a^{-1})^{-1} \bm{K}^T \bm{\Sigma}_\eps^{-1}$ and $\bm{A} = \bm{G} \bm{K}$ are called the retrieval gain matrix and the averaging kernel matrix, respectively \cite{rodgers2000}. The estimator for ${\theta} = \bm{h}^T \bm{x}$ is chosen to be the plug-in estimator $\hat{\theta} = \bm{h}^T \hat{\bm{x}}$.

To quantify the uncertainty of~$\theta$, we note that the posterior distribution on $\bm{x}$ induces a Gaussian posterior distribution on $\theta$ given by
\begin{equation}
\theta | \bm{y} \sim \mathcal{N}\big( \bm{h}^T (\bm{K}^T \bm{\Sigma}_\eps^{-1} \bm{K} + \bm{\Sigma}_a^{-1})^{-1} (\bm{K}^T \bm{\Sigma}_\eps^{-1} \bm{y} + \bm{\Sigma}_a^{-1} \bm{\mu}_a), \bm{h}^T (\bm{K}^T \bm{\Sigma}_\eps^{-1} \bm{K} + \bm{\Sigma}_a^{-1})^{-1} \bm{h} \big).
\end{equation}
A $(1-\alpha)$ central credible interval for $\theta$ is then given by
\begin{equation}
[\underline{\theta}, \overline{\theta}] = [\hat{\theta} - z_{1 - \alpha/2} \sigma, \hat{\theta} + z_{1 - \alpha/2} \sigma], \label{eq:XCO2_cred_int}
\end{equation}
where $\sigma^2 = \bm{h}^T (\bm{K}^T \bm{\Sigma}_\eps^{-1} \bm{K} + \bm{\Sigma}_a^{-1})^{-1} \bm{h}$ is the posterior variance of~$\theta$ and $\hat{\theta}$ the plug-in estimator of $\theta$, or equivalently the maximizer/mean of $p(\theta|\bm{y})$. The credible interval \eqref{eq:XCO2_cred_int} is used to quantify the uncertainty of $X_\mathrm{CO2}$ in the operational OCO-2 retrievals \cite{atbd2019}.

\subsection{Frequentist properties}

We describe in this section selected frequentist properties of the linearized operational retrieval method in order to compare its properties with those of our proposed method. It is straightforward to derive the following properties for the point estimator $\hat{\theta}$ and the credible interval $[\underline{\theta}, \overline{\theta}]$ given in Equation~\eqref{eq:XCO2_cred_int}:
\begin{itemize}
	
	\item \emph{Bias:} The bias of the estimator $\hat{\theta}$, denoted by $\mathrm{bias}(\hat{\theta})$, can be calculated as
	\begin{align}
	\mathrm{bias}(\hat{\theta}) &= \mathbb{E}_{\bm{\eps}}[\hat{\theta}] - \theta = \bm{h}^T (\mathbb{E}_{\bm{\eps}}[\hat{\bm{x}}] - \bm{x}) \label{eq:operational_bias} \\ &= \bm{h}^T (\bm{A}\bm{x} + (\bm{I}-\bm{A})\bm{\mu}_a - \bm{x}) = \bm{h}^T (\bm{A}-\bm{I})(\bm{x}-\bm{\mu}_a) = \bm{m}^T (\bm{x} - \bm{\mu}_a), \notag
	\end{align}
	where $\bm{m} = (\bm{A}^T-\bm{I})\bm{h} = \left( \bm{K}^T \bm{\Sigma}_\eps^{-1} \bm{K} (\bm{K}^T \bm{\Sigma}_\eps^{-1} \bm{K} + \bm{\Sigma}_a^{-1})^{-1} - \bm{I}\, \right) \bm{h}$ is a vector of bias multipliers. The bias depends on $\bm{x} - \bm{\mu}_a$, i.e., the difference between the true state $\bm{x}$ and the prior mean $\bm{\mu}_a$. Notice that the bias is 0 if and only if $\bm{x} = \bm{\mu}_a$ or $\bm{m} = \bm{0}$ or if the vector $\bm{x} - \bm{\mu}_a$ is orthogonal to $\bm{m}$. In other cases, depending on $\bm{x} - \bm{\mu}_a$ and how it interacts with the forward operator $\bm{K}$, the prior covariance $\bm{\Sigma}_a$, the noise covariance $\bm{\Sigma}_\eps$ and the functional~$\bm{h}$, there might be a positive or a negative~bias.
	
	\item \emph{Coverage:} As shown in Appendix~\ref{app:credibCoverage}, the frequentist coverage of the interval \eqref{eq:XCO2_cred_int} can be written down in closed form and is given by
	\begin{align}
		\mathbb{P}_{\bm{\eps}}(\theta \in [\underline{\theta}, \overline{\theta}])
		&= \Phi\left( \frac{\mathrm{bias}(\hat{\theta})}{\mathrm{se}(\hat{\theta})} + z_{1 - \alpha/2} \frac{\sigma}{\mathrm{se}(\hat{\theta})}\right) - \Phi\left( \frac{\mathrm{bias}(\hat{\theta})}{\mathrm{se}(\hat{\theta})} - z_{1 - \alpha/2} \frac{\sigma}{\mathrm{se}(\hat{\theta})}\right), \label{eq:operational_coverage}
	\end{align}
	where $\mathrm{se}(\hat{\theta}) = \sqrt{\mathrm{var}_{\bm{\varepsilon}}(\hat{\theta})}$ is the standard error of $\hat{\theta}$ and
	\begin{align}
	\mathrm{var}_{\bm{\varepsilon}}(\hat{\theta}) &= \mathrm{var}_{\bm{\varepsilon}}(\bm{h}^T \bm{G} \bm{\varepsilon}) = \bm{h}^T \bm{G} \bm{\Sigma}_\eps \bm{G}^T \bm{h} \label{eq:var_thetaHat} \\ &= \bm{h}^T (\bm{K}^T \bm{\Sigma}_\eps^{-1} \bm{K} + \bm{\Sigma}_a^{-1})^{-1} \bm{K}^T \bm{\Sigma}_\eps^{-1} \bm{K} (\bm{K}^T \bm{\Sigma}_\eps^{-1} \bm{K} + \bm{\Sigma}_a^{-1})^{-1} \bm{h} \notag
	\end{align} 
	is the variance of $\hat{\theta}$ computed with respect to the distribution of the noise $\bm{\eps}$. The coverage depends on $\bm{x}$ only through $\mathrm{bias}(\hat{\theta})$. It is an even function of $\mathrm{bias}(\hat{\theta})$ and the maximum is obtained with $\mathrm{bias}(\hat{\theta}) = 0$. In that case,
	\begin{align*}
	\mathbb{P}_{\bm{\eps}}(\theta \in [\underline{\theta}, \overline{\theta}]) &= \Phi\left(z_{1 - \alpha/2} \frac{\sigma}{\mathrm{se}(\hat{\theta})}\right) - \Phi\left(-z_{1 - \alpha/2} \frac{\sigma}{\mathrm{se}(\hat{\theta})}\right) \\
	&> \Phi\left(z_{1 - \alpha/2}\right) - \Phi\left(-z_{1 - \alpha/2}\right) = 1-\alpha,
	\end{align*}
	since $\sigma/\mathrm{se}(\hat{\theta}) > 1$.	In other words, the interval $[\underline{\theta}, \overline{\theta}]$ has overcoverage for $\mathrm{bias}(\hat{\theta}) = 0$. It is also easy to see that the coverage is a strictly decreasing function of $|\mathrm{bias}(\hat{\theta})|$. As $|\mathrm{bias}(\hat{\theta})|$ increases, the coverage eventually crosses the nominal value $(1-\alpha)$, followed by undercoverage. In the limit $|\mathrm{bias}(\hat{\theta})| \rightarrow \infty$, the coverage becomes zero.
	
	\item \emph{Length:} The interval $[\underline{\theta}, \overline{\theta}]$ has constant length given by $2 z_{1-\alpha/2} \sigma$.
	
	\item \emph{Comparison with standard error intervals:} A potential alternative for the credible interval \eqref{eq:XCO2_cred_int} is the frequentist standard error interval
	\begin{equation}
	[\underline{\theta}, \overline{\theta}] = [\hat{\theta} - z_{1 - \alpha/2} \,\mathrm{se}(\hat{\theta}), \hat{\theta} + z_{1 - \alpha/2} \,\mathrm{se}(\hat{\theta})]. \label{eq:stdErrInt}
	\end{equation}
	It is easy to show that the credible interval \eqref{eq:XCO2_cred_int} is always longer than the standard error interval \eqref{eq:stdErrInt}. This extra length can be understood as an attempt to inflate the uncertainties to account for the bias~\eqref{eq:operational_bias}; see Section~6.4 in \cite{ruppert2003} and the references therein. It follows that the coverage of the credible interval \eqref{eq:XCO2_cred_int} is greater than that of the standard error interval \eqref{eq:stdErrInt}, which undercovers whenever $\mathrm{bias}(\hat{\theta}) \neq 0$ \cite{kuusela2016}.
\end{itemize}

\subsection{Commentary}

The operational retrieval method is based on the well-established Bayesian framework where the observed data are combined with the prior distribution to obtain inferences in the form of the posterior distribution. The operational inferences should therefore be interpreted as representing a Bayesian degree of belief about $\theta$. However, a user of the retrieval method may also be interested in frequentist inference of $\theta$ and the above analysis shows that the operational method can be miscalibrated if used for frequentist inference. As is well known, the performance of Bayesian methods can depend critically on the choice of the prior distribution, and the same is true for the frequentist properties of the operational method. For example, the point estimator $\hat{\theta}$ would be unbiased if the prior mean was chosen to be equal to the true state, i.e., $\bm{\mu}_a = \bm{x}$, but this is unlikely in practice as it would require knowing beforehand what the value of $\bm{x}$ is. (The bias is also small if $\bm{x} - \bm{\mu}_a$ is nearly orthogonal to $\bm{m}$ but this is equally unlikely to hold true.) At least some amount of frequentist bias is therefore always present, with the potential for arbitrarily large biases depending on how much the prior mean deviates from the true state. Since the frequentist coverage of the intervals depends on the bias, this can result in wildly varying coverage performance. For small biases, the intervals overcover, i.e., $\mathbb{P}_{\bm{\eps}}(\theta \in [\underline{\theta}, \overline{\theta}]) > 1-\alpha$, while for large biases the intervals undercover, i.e., $\mathbb{P}_{\bm{\eps}}(\theta \in [\underline{\theta}, \overline{\theta}]) < 1-\alpha$. Irrespective of which of these two cases dominates, the intervals are bound to have some degree of frequentist miscalibration since it is unlikely that there would always be just the right amount of bias for nominal coverage. Since it is impossible to judge the coverage of the intervals without knowing the true $\bm{x}$, it is not possible to tell for real soundings if a given interval is well-calibrated or not. Ideally, roughly $100\times(1-\alpha)\%$ of soundings from a given OCO-2 orbit would cover their true $X_\mathrm{CO2}$ values. However, this discussion shows that, for the current retrieval method, this fraction can be much smaller or much larger. Furthermore, in the real atmosphere, the nearby states $\bm{x}$ are spatially and temporally correlated. Since the bias depends on $\bm{x} - \bm{\mu}_a$, this means that the biases, and therefore also the coverage values, are spatially and temporally correlated, which may lead to misleading frequentist inferences over extended spatial regions or temporal periods. These effects are analyzed in greater detail using a simulated example scenario in Section~\ref{sec:numerical}. It is also worth noting that these suboptimal frequentist properties of the operational method are not unexpected as a Bayesian method is not necessarily designed to have good frequentist properties. Indeed, the above issues are not necessarily problematic when seen from the Bayesian perspective. It is also possible to modify a Bayesian procedure to improve its frequentist properties \cite{Bayarri2004,Berger2006,Kass1996}; however, in this work we focus on the operational retrieval method as it is currently implemented in OCO-2.

\section{Numerical results}
\label{sec:numerical}

\subsection{Experiment setup}
\label{sec:numerical_setup}

\subsubsection{Forward model and weight vector specifics}
\label{sec:numerical_model}

The starting point for our forward model is the OCO-2 surrogate model developed by Hobbs et al \cite{hobbs2017}. The surrogate model is a computationally efficient approximation to the OCO-2 Full Physics forward model \cite{atbd2019}. Similar to the full model, it involves a nonlinear mapping from the state vector $\bm{x}$ to the radiances~$\bm{y}$, but is much faster to evaluate. The surrogate model also makes certain simplifications to the full OCO-2 state vector. As described in Section~\ref{sec:mathematical_model}, we make a further approximation by linearizing the surrogate model, which leads to the linear model in Equation~\eqref{eq:linear_forward_model}. The linearization is done around the generative process mean $\bm{\mu}_{\bm{x}}$; see Section~\ref{subsec:numerical_data}.

The state vector $\bm{x}$ in the surrogate model has 39 elements ($p = 39$) of which the first 20 correspond to the vertical CO$_2$ profile and the remaining 19 are nuisance variables related to surface pressure ($x_{21}$), surface albedo ($x_{22},\ldots,x_{27}$) and atmospheric aerosol concentrations ($x_{28},\ldots,x_{39}$). A detailed description of these variables is given in the Supplement \cite{Patil2021Supp}; see also \cite{hobbs2017}. These variables suffice in order to capture, to a good approximation, the relation between the atmospheric CO$_2$ profile $x_1,\ldots,x_{20}$ and the observed radiances $\bm{y}$ \cite{hobbs2017}.

In addition to the state vector $\bm{x}$, the forward operator depends on additional parameters, most notably on the solar and satellite viewing geometries, which are assumed to be known during the retrieval. In our case, the forward model is evaluated for an OCO-2 orbit that took place in October 2015 near the Lamont, Oklahoma, Total Carbon Column Observing Network (TCCON) site (36.604$^\circ$N, 97.486$^\circ$W). The satellite is in the nadir observing mode, i.e., pointed toward the ground directly underneath its orbit.

We can investigate the ill-posedness of the CO$_2$ retrieval problem by studying the singular values of the linearized forward operator represented by the $3048 \times 39$ matrix $\bm{K}$. The singular values, shown in Figure~S2 in the Supplement \cite{Patil2021Supp}, decay exponentially indicating that the retrieval problem is severely ill-posed \cite{hansen2005}. The smallest singular value deviates from the exponential decay which we take to indicate that $\bm{K}$ is rank deficient with rank 38. Hence, there is a one-dimensional null space. The condition number (the ratio of the largest to the smallest (numerically) non-zero singular value) is $3.62 \times 10^{12}$, consistent with a severely ill-posed problem.

The ultimate quantity of interest in the retrieval problem is the column-averaged CO$_2$ dry-air mole fraction $X_\mathrm{CO2} = \tp{\bm{h}}\bm{x}$, where $\bm{h}$ is a weight vector derived in \cite{ODell2012}; see also~\cite{atbd2019}. Since $X_\mathrm{CO2}$ only involves the CO$_2$ profile, the nuisance variables get weight zero, i.e., $h_{21} = \cdots = h_{39} = 0$. The remaining weights are strictly positive and sum to one, $\sum_{i=1}^{20} h_i = 1$, so statistically $X_\mathrm{CO2}$ is a weighted average of the CO$_2$ concentrations $x_1,\ldots,x_{20}$. In the Full Physics retrievals, the weights $h_i$ have a slight dependence on the nuisance variables, but in the surrogate model the weights do not depend on the state vector. In practice, the surrogate model weights are almost constant for the intermediate pressure levels, while the weights for the boundary levels are approximately half of that value.

\subsubsection{Data generation}
\label{subsec:numerical_data}

Our investigations require a realistic generative model from which synthetic states and observations can be simulated. A suitable multivariate distribution for the state vector $\bm{x}$, as well as a model for the spatial dependence among state vectors in a small spatial region, was developed in \cite{hobbs2021}. Briefly, the approach uses actual retrieved state vectors near Lamont, OK, during the month of October 2015. This collection is part of the OCO-2 Level 2 diagnostic data products, available at the NASA Goddard Earth Science Data and Information Services Center (GES DISC, \url{https://disc.gsfc.nasa.gov/OCO-2}). These are combined with a simulation-based assessment of the retrieval error properties to estimate the state vector mean $\bm{\mu}_{\bm{x}}$ and the single-sounding covariance $\bm{\Sigma}_{\bm{x}}$ for this location and time.

Synthetic data are then generated through the following steps:

\begin{enumerate}
	\item \emph{State vector generation for a single sounding:} $\bm{x} \sim \mathcal{N}(\bm{\mu}_{\bm{x}},\bm{\Sigma}_{\bm{x}})$, where the parameters of the multivariate Normal distribution were estimated from OCO\nobreakdash-2 data as noted above.
	
	\item \emph{State vector generation for grid sounding:} We also simulate states $\bm{x} (\bm{s}_i)$ on a grid of $i = 1, \ldots, 64$ locations within an OCO-2 orbit. Following \cite{hobbs2021}, we assume that this spatial process $\bm{x}(\cdot) \sim \mathrm{GP}(\bm{\mu}(\cdot),\bm{C}(\cdot,\cdot))$ is 
    a multivariate Gaussian process with a spatially constant mean function $\bm{\mu}(\cdot) = \bm{\mu}_{\bm{x}}$ and cross-covariance function $\bm{C}(\cdot,\cdot)$ defined as $C_{kl} (\bm{s}_i,\bm{s}_j) = \cov{x_k (\bm{s}_i), x_l (\bm{s}_j)} = \Sigma_{\bm{x},kl}\, \mathcal{M}_{kl}\left( \|\bm{s}_i-\bm{s}_j\| \right)$, where $\mathcal{M}_{kl}$ is a Mat\'ern-type correlation function \cite{stein1999}. The parameters of the correlation function vary across $k$ and $l$ in a way that guarantees positive definiteness and were estimated from the above collection of OCO-2 retrieved state vectors~\cite{hobbs2021}.
	
	\item \emph{Noise generation:} $\bm{\eps} \sim \mathcal{N}({\bm{0}, \bm{\Sigma}_{\bm{\eps}}})$.
    The OCO-2 radiances are fundamentally photon counts in the detectors so these measurements have Poisson-like behavior. The noise can nevertheless be approximated well using an additive Gaussian noise term with zero mean and variance 
    proportional to the mean signal. Following \cite{hobbs2017} and \cite{lamminpaa2019}, we let $\bm{\Sigma}_{\bm{\eps}}$ be diagonal with elements $\var{\varepsilon_{j}} = c_{b(j)} F_{j} (\bm{\mu}_{\bm{x}})$, where $b: \{1,\ldots,3048\} \rightarrow \{1,2,3\},\ j \mapsto b(j)$ indicates the spectral band (O$_2$, weak CO$_2$, strong CO$_2$) of $j$, $c_i$ are band-specific constants and $F_{j}(\cdot)$ is the $j$th element of the forward operator output. In the actual satellite, the noise model is somewhat more complicated, but its properties are nevertheless well-understood \cite{Crisp2021}. The $\bm{\varepsilon}$ realizations are i.i.d.\ both across repetitions of the experiment for a fixed state $\bm{x}$ and over the different spatial sounding locations.
	
	\item \emph{Radiance observation:} $\bm{y} = \bm{K} \bm{x} + \bm{\eps}$, where $\bm{x}$ and $\bm{\eps}$ are given by the previous steps and the matrix $\bm{K}$ results from linearizing the forward operator $\bm{F}$ about the true mean~$\bm{\mu}_{\bm{x}}$.
	
\end{enumerate}
In addition, the operational procedure posits a prior distribution on the state $\bm{x}$, which is given by $\bm{x} \sim \mathcal{N}(\bm{\mu}_a, \bm{\Sigma}_a)$. We use the prior mean $\bm{\mu}_a$ and prior covariance $\bm{\Sigma}_a$ derived from the OCO-2 operational prior near Lamont, OK, in October 2015. For OCO-2, the prior mean $\bm{\mu}_a$ varies in space and time but is dependent in part on climatology and expert knowledge, while $\bm{\Sigma}_a$ is the same for all retrievals.

An important point to highlight is that $\bm{\mu}_a \neq \bm{\mu}_{\bm{x}}$ and $\bm{\Sigma}_a \neq \bm{\Sigma}_{\bm{x}}$.  Therefore, the true conditions, represented through $\bm{\mu}_{\bm{x}}$ and $\bm{\Sigma}_{\bm{x}}$ in our simulations, will be different from the prior mean and covariance. This misspecification is a real challenge for the operational retrievals and a source of bias \cite{nguyen2019}. The prior model and the generative model are visualized and compared in detail in the Supplement \cite{Patil2021Supp} which also contains a visualization of the spatial dependence structure of the state vectors.

\subsubsection{Constraints} \label{sec:constraints}

In the proposed frequentist procedure, we impose non-negativity constraints on certain elements of the state vector $\bm{x}$. Since elements $x_1,\ldots,x_{20}$ are CO$_2$ concentrations, they need to be non-negative by definition. Thus, we impose the constraint $x_i \geq 0$ for $i = 1, \dots, 20$. The same argument applies to surface pressure so we also include the constraint  $x_{21} \geq 0$. The rest of the state vector elements are left unconstrained.

Since albedo is a fraction between 0 and 1, this implies in principle linear inequality constraints for the albedo variables $x_{22},\dots,x_{27}$. We experimented with adding these constraints but found that that made little difference in the results while causing some extra computational overhead. We therefore decided to leave these variables unconstrained. The aerosol variables $x_{28},\dots,x_{39}$ are parameterized in the surrogate model in such a way that there are no trivial constraints that could be imposed on those variables.

\subsection{Single sounding results}
\label{sec:numerical_single}

\subsubsection{Distribution of bias of the operational method} \label{sec:bias_distribution}

\begin{figure}[t]
	\centering
	\subfigure[Bias multipliers]{
	\includegraphics[height=0.365\columnwidth]{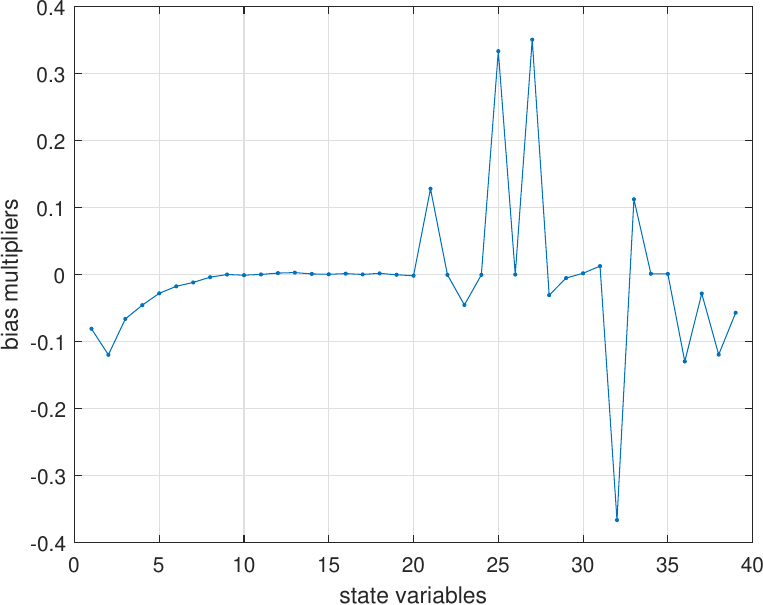}
	\label{fig:bias_multipliers}}
	\subfigure[Bias components]{
	\includegraphics[height=0.365\columnwidth]{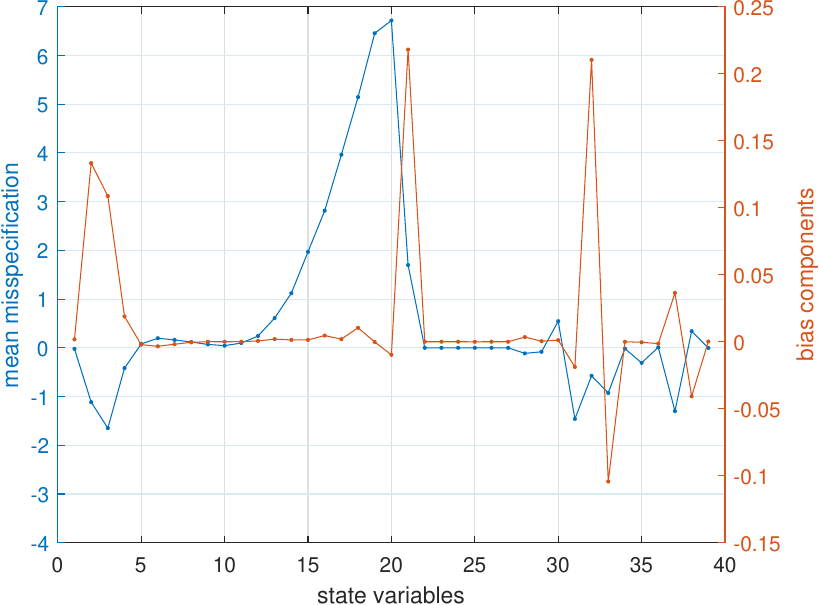}
	\label{fig:state_difference_bias_components_mean_misspecification}}
	\caption{Figure \subref{fig:bias_multipliers} illustrates the bias multiplier vector $\bm{m}$, while Figure \subref{fig:state_difference_bias_components_mean_misspecification} shows the corresponding bias components $m_i (\mu_{\bm{x},i}-\mu_{a,i})$ for the misspecified means.}
\end{figure}

Since the linearized operational method is based on a linear estimator $\hat{\theta}$, we can write down $\mathrm{bias}(\hat{\theta}) = \mathbb{E}_{\bm{\eps}}[\hat{\theta}] - \theta$ in closed form for a given $\bm{x}$. This is done in Equation~\eqref{eq:operational_bias}, which shows that the bias is given by the inner product of the bias multiplier vector $\bm{m}$ and the prior mean misspecification $\bm{x}-\bm{\mu}_a$. For a given $\bm{x}$ sampled from the generative model, there will therefore always be a nonzero bias whose size depends on the details of the prior misspecification for that particular $\bm{x}$. To understand this interaction better, we show the bias multiplier vector $\bm{m}$ in Figure \ref{fig:bias_multipliers} for our particular retrieval setup. This highlights the role of the nuisance variables $x_{21},\ldots,x_{39}$ in dictating the size of the bias. Notice that the bias multiplier $\bm{m}$ depends on the prior covariance $\bm{\Sigma}_a$ but not on the prior mean $\bm{\mu}_a$. Hence this can be seen as a way of decoupling the contribution of the prior mean on the bias from that of the prior covariance. It is also worth noting that here the bias is entirely caused by the regularization in the prior since we generate the data using the same linear forward model $\bm{K}$ that we use in the inversion; in real-life retrievals, there might be an additional component in the bias from the nonlinearity of the forward operator.

\begin{figure}[t]
	\centering
	\subfigure[Bias distribution]{
	\includegraphics[height=0.36\columnwidth]{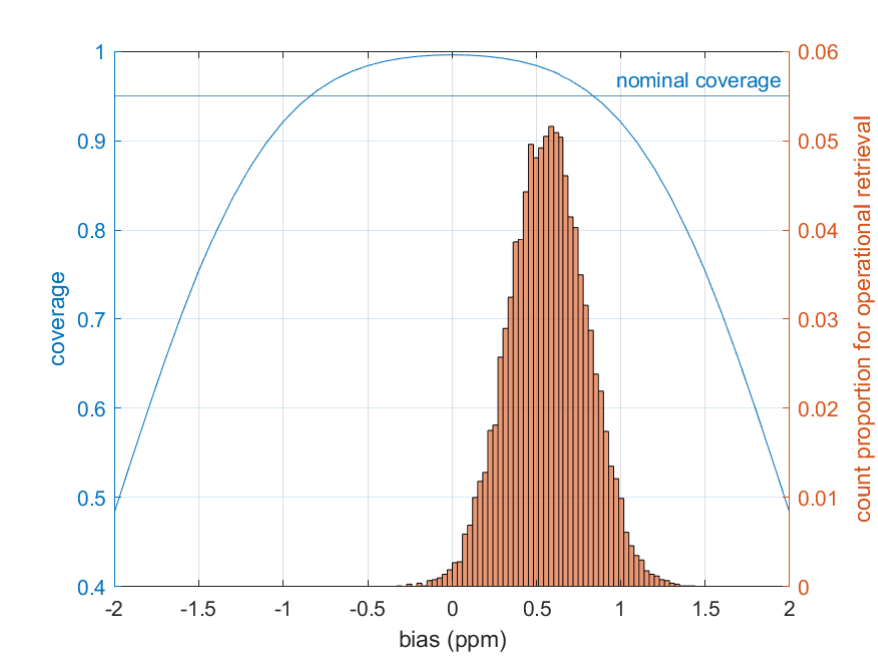}
	\label{fig:operational_bias_coverage_singlesounding_seed255}}
	\subfigure[Coverage distribution]{
	\includegraphics[height=0.36\columnwidth]{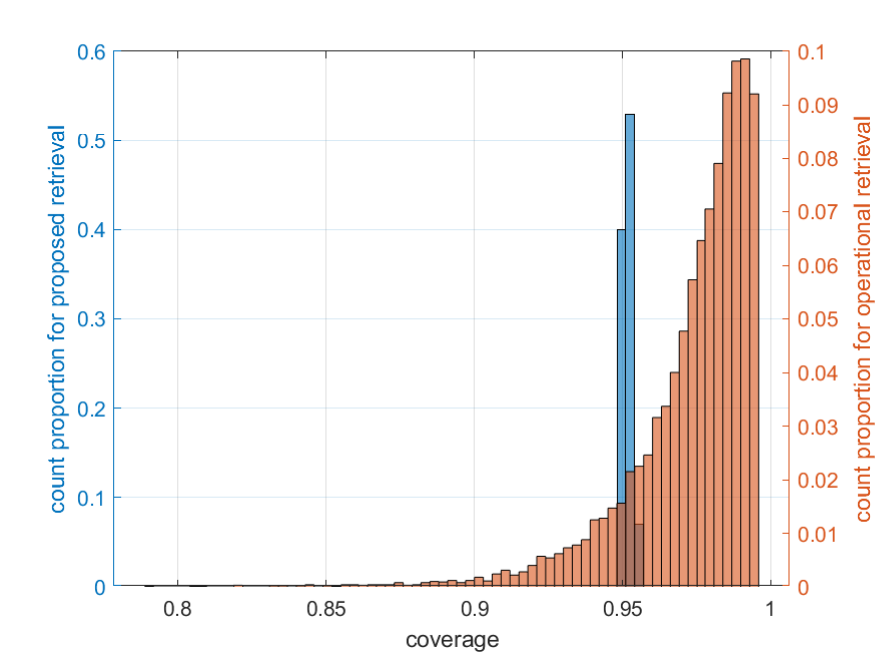}
	\label{fig:operational_vs_proposed_coverage_singlesounding_seed255}}
	\caption{Figure \subref{fig:operational_bias_coverage_singlesounding_seed255} shows the coverage as a function of bias (blue line) for the operational procedure and the corresponding histogram of operational retrieval bias. Figure \subref{fig:operational_vs_proposed_coverage_singlesounding_seed255} shows a histogram (in orange color) of the operational retrieval coverage for $95\%$ intervals. Also shown is a histogram (in blue color) of empirical coverage for the proposed frequentist uncertainty quantification method.}
\end{figure}

Assuming that $\bm{x} \sim \mathcal{N}(\bm{\mu}_{\bm{x}},\bm{\Sigma}_{\bm{x}})$ gives a realistic distribution of $\bm{x}$'s for repeated satellite overpasses, we can also derive the distribution of the bias over repeated $\bm{x}$ realizations. In our particular case, we have $\mathrm{bias}(\hat{\theta}) \sim \mathcal{N}(0.5714, 0.0533)$. This distribution is illustrated in Figure~\ref{fig:operational_bias_coverage_singlesounding_seed255} showing the bias for 10\,000 instances of $\bm{x}$ from the generative model. This shows that the biases are typically positive with a fair amount of spread around the central value. Negative biases and biases larger than $1.2\ \mathrm{ppm}$ are rare, at least in this particular setup for the retrieval problem.

We have that $\mathbb{E}_{\bm{x}}[\mathrm{bias}(\hat{\theta})] = \bm{m}^T (\bm{\mu}_{\bm{x}} - \bm{\mu}_a)$, which corresponds to the bias expressions given in \cite{nguyen2019}. Hence, the distribution of $\mathrm{bias}(\hat{\theta})$ has mean zero if and only if $\bm{\mu}_a = \bm{\mu}_{\bm{x}}$ or if $\bm{\mu}_{\bm{x}} - \bm{\mu}_a$ is orthogonal to $\bm{m}$. Even in those cases, $\mathrm{bias}(\hat{\theta})$ would still have a spread around zero so individual retrievals may be positively or negatively biased. In the more realistic case where $\bm{m}^T (\bm{\mu}_{\bm{x}} - \bm{\mu}_a) \neq 0$, the biases are either predominantly positive or negative depending on the details of the prior misspecification. Figure~\ref{fig:state_difference_bias_components_mean_misspecification} shows a breakdown of the contribution of each state variable to the mean bias of 0.5714 in our particular setup. The figure visualizes the mean misspecification $\bm{\mu}_{\bm{x}} - \bm{\mu}_a$ and the individual terms $m_i (\mu_{\bm{x},i}-\mu_{a,i})$ contributing to the mean bias. It enables us to conclude that the positive biases are primarily caused by the misspecification of the surface pressure variable $x_{21}$, the aerosol variable $x_{32}$ and the upper portion of the CO$_2$ profile, all of which contribute positively to the mean bias. The large misspecification of the lower portion of the CO$_2$ profile, on the other hand, makes negligible contribution to the bias due to the small bias multipliers of those variables.

\subsubsection{Coverage and length of the operational and proposed intervals} \label{sec:distOperCover}

The frequentist coverage of the operational method for a particular $\bm{x}$ defined as $ \mathbb{P}_{\bm{\eps}}(\theta \in [\underline{\theta}, \overline{\theta}])$ can be calculated using Equation~\eqref{eq:operational_coverage}. The coverage depends on $\bm{x}$ only through $\mathrm{bias}(\hat{\theta})$. To understand the nature of this dependence, we plot in Figure~\ref{fig:operational_bias_coverage_singlesounding_seed255} the coverage of $95\%$ intervals as a function of the bias for our particular retrieval setup. We observe that for $|\mathrm{bias}(\hat{\theta})| < 0.84 \text{ ppm}$ the intervals overcover, while for $|\mathrm{bias}(\hat{\theta})| > 0.84 \text{ ppm}$ the intervals undercover, with the coverage dropping sharply for biases larger than 1 ppm in absolute value. Since our biases are predominantly positive (Figure~\ref{fig:operational_bias_coverage_singlesounding_seed255}), we are mostly going to observe the subrange of coverages corresponding to $\mathrm{bias}(\hat{\theta}) \in [-0.3 \text{ ppm}, 1.4 \text{ ppm}]$.

It is difficult to explicitly write down the distribution of the coverage corresponding to the assumed distribution of $\bm{x}$, but we evaluate the coverage distribution numerically in Figure~\ref{fig:operational_vs_proposed_coverage_singlesounding_seed255} for 10\,000 state vectors and $95\%$ intervals. We see that the operational intervals are poorly calibrated in terms of their frequentist coverage. For most $\bm{x}$ realizations, the intervals have overcoverage. However, we also note that the coverage distribution is heavily left-skewed toward values below the nominal $95\%$ coverage. In particular, for 12.03\% of the $\bm{x}$ realizations, the intervals have undercoverage. The smallest coverage is $79.0\%$, and this could drop even lower depending on the $\bm{x}$ realization.

Such coverage behavior is inherent to the operational retrieval method because of the bias induced by the regularizing prior. This leads to the somewhat paradoxical conclusion that, if interpreted as frequentist confidence intervals, the operational intervals are too long for most $\bm{x}$'s, while for roughly $12\%$ of the $\bm{x}$'s, the intervals are too short. Unfortunately, there is no easy way of telling when the intervals are too long or too short, so it is not possible to adaptively recalibrate their length.

The proposed frequentist direct retrieval method, on the other hand, has fundamentally different behavior. For this method, it is not straightforward to write down a closed-form expression for the coverage, but we can nevertheless evaluate it empirically. Here we evaluate the empirical coverage of 95\% intervals using 10\,000 realizations of the noise~$\bm{\eps}$. This is repeated for 100 realizations of $\x$ from the generative model to study the distribution of the coverage values. The results are shown in Figure~\ref{fig:operational_vs_proposed_coverage_singlesounding_seed255}. We find that the proposed method is well-calibrated across all considered $\x$ instances. The coverage peaks at slightly above 95\%, with very little spread around that value. For some $\x$, the intervals have a small amount of overcoverage, but this is very minor in comparison to the operational method.

To further compare the two methods, we pick 10 instances of $\bm{x}$ corresponding to 10 different coverage values for the operational method ranging from the minimum operational coverage to the maximum in Figure~\ref{fig:operational_vs_proposed_coverage_singlesounding_seed255}. Table~\ref{tab:coverage_comparison_single_sounding} compares the 95\% intervals for the two methods for each of these 10 state vectors. We observe that while the coverage of the operational method can vary between substantial undercoverage and major overcoverage, the proposed method consistently achieves nearly nominal coverage irrespective of the $\bm{x}$ realization.

\begin{table}[t] \label{tab:coverage_comparison_single_sounding}
	\centering
	\caption{\noindent Comparison of coverage and interval length (in ppm) between the operational and proposed uncertainty quantification methods for 10 state vector $\bm{x}$ realizations chosen uniformly between the minimum and maximum coverage for the operational method in Figure~\ref{fig:operational_vs_proposed_coverage_singlesounding_seed255}. The target coverage in each case is 95\%. Also shown are the bias of the operational point estimates and the standard deviation of the length of the proposed intervals (both in ppm). The proposed method is not based on a point estimator, so we do not report a bias value for it.}
	\begin{tabular}{lllllll}
		\toprule
		$\bm{x}$ & operational & operational & operational & proposed & proposed & proposed \\
		realization & bias & coverage & length & coverage & avg.\ length & length s.d. \\
		\midrule
		1 & 1.4173 & 0.7899 & 3.94 & 0.9515 & 11.20 & 0.29 \\ 
		2 & 1.3707 & 0.8090 & 3.94 & 0.9511 & 11.20 & 0.28  \\ 
		3 & 1.2986 & 0.8363 & 3.94 & 0.9510 & 11.20 & 0.29 \\ 
		4 & 1.2357 & 0.8579 & 3.94 & 0.9515 & 11.20 & 0.28 \\ 
		5 & 1.1590 & 0.8816 & 3.94 & 0.9513 & 11.20 & 0.28 \\ 
		6 & 1.0747 & 0.9042 & 3.94 & 0.9512 & 11.21 & 0.27 \\ 
		7 & 0.9721 & 0.9272 & 3.94 & 0.9515 & 11.20 & 0.29 \\ 
		8 & 0.8420 & 0.9500 & 3.94 & 0.9513 & 11.19 & 0.31 \\ 
		9 & 0.6477 & 0.9730 & 3.94 & 0.9508 & 11.19 & 0.32 \\ 
		10 & 0.0001  & 0.9959 & 3.94 & 0.9502 & 11.18 & 0.35 \\ 
		\bottomrule
	\end{tabular}
\end{table}

The two approaches also have different behaviors in terms of their interval lengths. The operational intervals have constant length $2z_{1-\alpha/2}\sigma$, where $\sigma$ is the posterior standard deviation of $\theta$ that does not depend on the data $\y$. In our case, $\sigma = \nobreak 1.0051\text{ ppm}$, so the operational intervals have constant length of 3.94~ppm at $95\%$ confidence level. It is worth noting that $\mathrm{se}(\hat{\theta}) = 0.6856\text{ ppm}$. Hence, the operational intervals derived from the posterior of $\theta$ are almost 50\% longer than what standard error intervals would be. This extra length gives the operational intervals some, but not enough, protection against undercoverage.

The proposed intervals, on the other hand, have data-dependent length. We report in Table~\ref{tab:coverage_comparison_single_sounding} the average lengths and length standard deviations for these intervals across different $\bm{\eps}$ realizations for each fixed state vector $\bm{x}$. We observe that the interval lengths indeed vary across noise realizations with a coefficient of variation (ratio of standard deviation to average length) of about $3\%$. However, the average lengths are almost constant across different $\bm{x}$'s. We therefore conclude that, while the proposed intervals have variable length, their average length does not seem to depend much on the true state $\bm{x}$.

The proposed intervals are longer than the operational ones, but in exchange seem to provide the desired coverage irrespective of the $\bm{x}$ realization. We can gain further insight into the behavior of the two constructions by visualizing some illustrative realizations of the state vectors and the associated intervals produced by the two methods. This is done in Section S3 in the Supplement \cite{Patil2021Supp}.

\subsection{Coverage over a spatial region}
\label{sec:numerical_grid}

In this section, we investigate the performance of the operational method over a spatial grid of $8 \times 8$ soundings near Lamont, OK. The size of the region is approximately 8~km in the cross-track direction and 16~km in the along-track direction. We generate spatially correlated state vectors $\bm{x}(\bm{s}_i)$ and expected radiances $\bm{K}\bm{x}(\bm{s}_i)$ over the grid as described in Section~\ref{subsec:numerical_data}. We then investigate the bias of the operational point estimates and the pointwise coverage of the operational 95\% intervals over the grid. To do this, we can simply use the closed-form expressions \eqref{eq:operational_bias} and \eqref{eq:operational_coverage} at each sounding location. Since both the bias and the coverage of the operational method are functions of the state vectors $\bm{x}(\bm{s}_i)$, these properties inherit the spatial dependence between the state vectors and will hence exhibit spatially correlated patterns.

\begin{figure}[t]
	\centering
	
	\subfigure[Operational coverage]{
		\includegraphics[height=0.345\columnwidth]{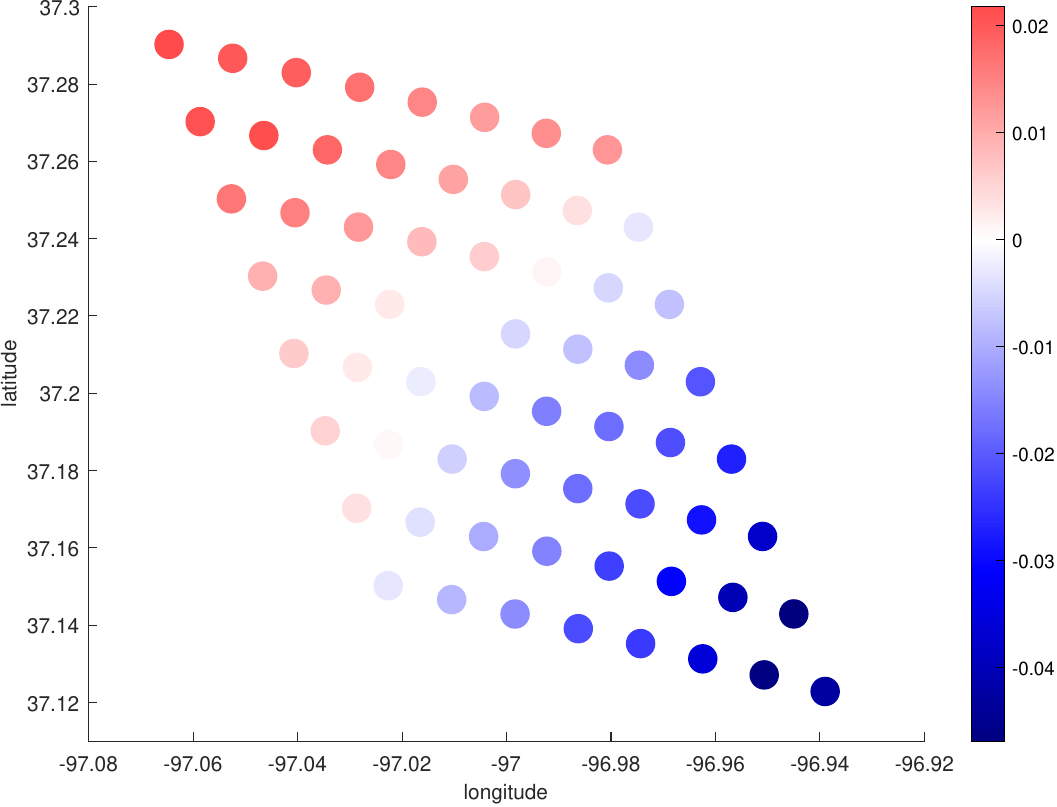}
		\label{fig:operational_grid_coverage_763}}
	\subfigure[Operational bias]{\includegraphics[height=0.345\columnwidth]{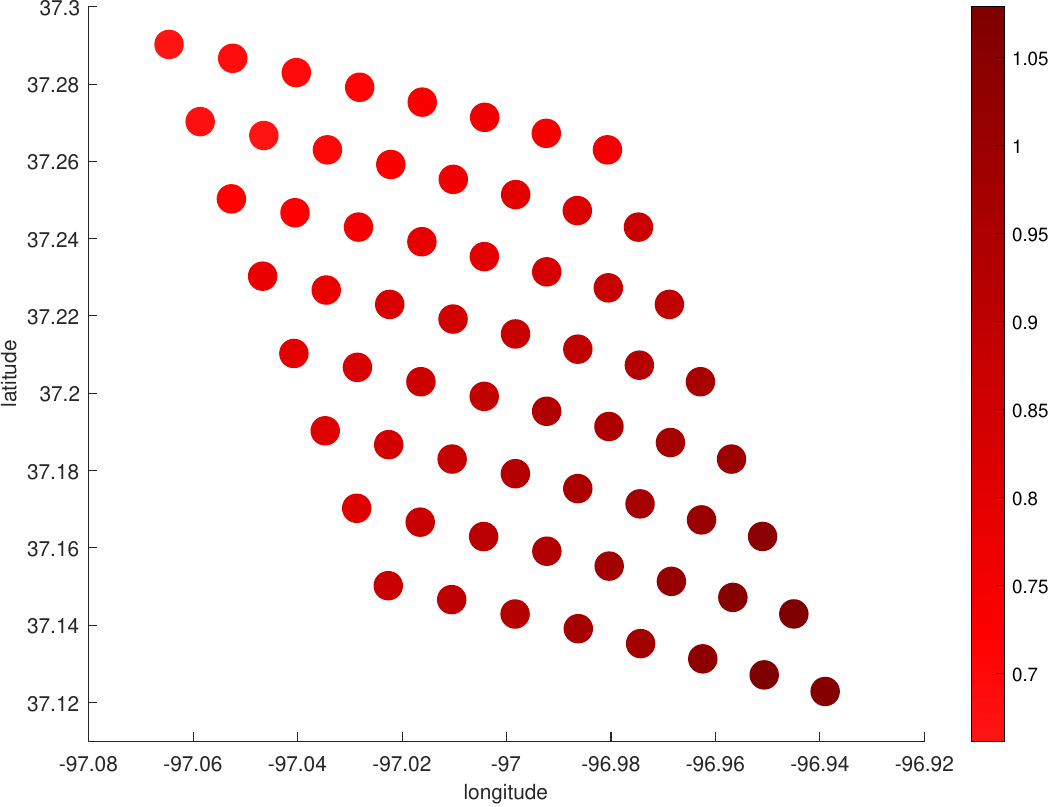}
		\label{fig:operational_grid_bias_763}}
	
	\caption{Operational retrieval over a grid of $8 \times 8$ soundings for an instance where both undercoverage and overcoverage are present. Figure~\subref{fig:operational_grid_coverage_763} shows the spatial coverage pattern relative to the nominal 95\% in units of probability (i.e., -0.03, for example, corresponds to coverage 0.92, instead of the nominal 0.95). The fraction of soundings below nominal coverage is 0.55. Figure~\subref{fig:operational_grid_bias_763} shows the corresponding bias pattern in ppm. }
	\label{fig:operational_grid_763}
\end{figure}

The observed patterns depend on the specific $\{\bm{x}(\bm{s}_i) : i=1,\ldots,64\}$ realization over the grid. Figure~\ref{fig:operational_grid_coverage_763} shows the coverage pattern for a case in which the coverage systematically changes from overcoverage to undercoverage when moving from the northwest corner of the grid to the southeast corner. The reason for this can be seen from the bias pattern in Figure~\ref{fig:operational_grid_bias_763} which shows that the overall positive bias has a systematic gradient across the region so that the bias is larger in the southeast corner and smaller in the northwest corner, which then affects the coverage as described in Equation~\eqref{eq:operational_coverage}. It is also possible to observe undercoverage over the entire grid. Figure~S9 in the Supplement \cite{Patil2021Supp} shows the coverage and bias patterns for a case where the state vector realizations are such that all 64 intervals across the region have coverage below the nominal value due to a systematic large positive bias throughout the region.

These results illustrate one of the challenges of the operational retrieval method in that there can be entire regions with undercoverage or overcoverage. For example, in the case of Figure~S9, all intervals across the region are systematically offset toward too large $X_\mathrm{CO2}$ values which causes their lower bounds to miss the corresponding true $X_\mathrm{CO2}$ values more often than they should. There is a risk that such patterns could be mistaken as CO$_2$ flux signals. As such, these observations may have important implications for carbon flux estimates; see Section~\ref{sec:conclusion}. The coverage patterns shown here for the operational method are in contrast with the behavior of the proposed frequentist method which does not exhibit systematic spatially correlated~miscalibration.

\section{Variable importance and effect of additional constraints on interval length}
\label{sec:extension}

As demonstrated in the previous section, the proposed frequentist method has good coverage performance, but the intervals are longer than in the operational retrieval method. In this section, we consider different variants of the proposed method to improve the interval length.

It is worth noting that the underlying inference problem is defined by (i) the forward operator $\bm{K}$, (ii) the functional of interest parameterized by the weight vector $\bm{h}$, (iii) the amount of noise in the problem controlled by the covariance $\bm{\Sigma}_{\bm{\varepsilon}}$, and (iv) the constraint set~$C$. For a given sounding and quantity of interest, we cannot change $\bm{K}$, $\bm{\Sigma}_{\bm{\varepsilon}}$ or $\bm{h}$, but we can potentially alter $C$. We therefore investigate how changes in $C$ in the form of additional constraints affect the length of the proposed intervals.

\subsection{Effect of individual nuisance variables}

We have so far only used trivial positivity constraints on certain state vector elements; see Section~\ref{sec:constraints}. However, if additional information was available to further constrain the state vector---for example, one could imagine that observational data from other sources tell us that, for some $i$, $x_i \in [\underline{x}_i, \overline{x}_i]$ with high probability---then including those constraints should result in shorter intervals for $X_\mathrm{CO2}$. 

\begin{figure}[h]
	\centering
	\subfigure[Nuisance variable importance]{
		\includegraphics[height=0.37\columnwidth]{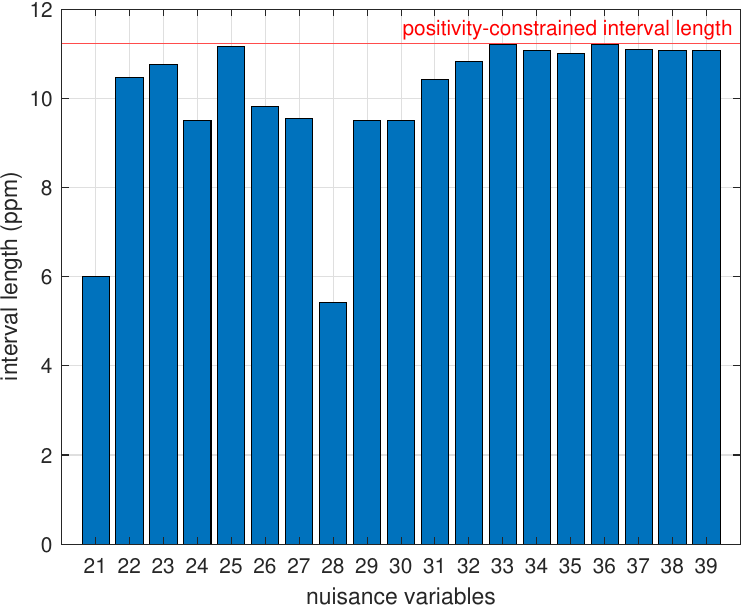}
		\label{fig:interval_length_one_fixed_nuance_seed436}}
	\subfigure[Pressure importance]{
		\includegraphics[height=0.37\columnwidth]{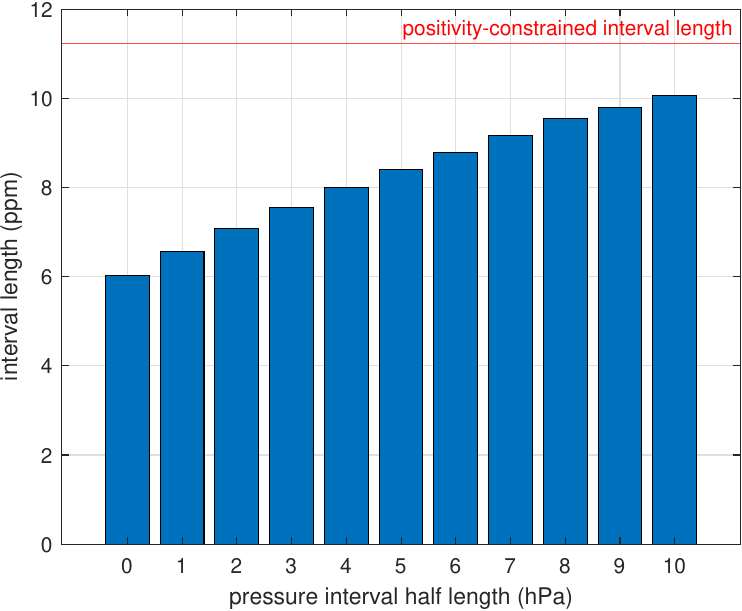}
		\label{fig:interval_length_varying_pressure_seed129}}
	\caption{Figure \subref{fig:interval_length_one_fixed_nuance_seed436} shows average $X_\mathrm{CO2}$ interval lengths at 95\% confidence level when constraining one nuisance variable at a time to its true value. Figure \subref{fig:interval_length_varying_pressure_seed129} shows average $X_\mathrm{CO2}$ interval lengths at 95\% confidence level for varying degrees of deterministic constraints on the surface pressure variable $x_{21}$. In both figures, the horizontal red line shows the interval length when only trivial positivity constraints are used.}
\end{figure}

To investigate what impact this would have, we start by considering how each nuisance variable $x_{21},\ldots,x_{39}$ affects the final $X_\mathrm{CO2}$ interval length. Figure \ref{fig:interval_length_one_fixed_nuance_seed436} shows the average interval lengths for the proposed method when one of the nuisance variables is assumed to be known. We can incorporate this assumption by using constraints of the form $x_{i,\textrm{true}} \le x_i \le x_{i,\textrm{true}}$ for one $x_i$ at a time, in addition to the previously used positivity constraints. As expected, the interval lengths are smaller than the interval length without any additional information. In particular, variables $x_{21}$ (surface pressure) and $x_{28}$ (log-AOD for the first composite aerosol type; see Supplement \cite{Patil2021Supp}) have the greatest impact on the interval length. Therefore additional constraints on these two variables could be particularly helpful in reducing the interval length. Since it is not immediately clear what observational constraints might be available for $x_{28}$, we will in the following focus on constraints for the pressure variable $x_{21}$.

\subsection{Deterministic pressure constraints}

We analyze the effect of various degrees of deterministic constraints on the pressure variable in Figure~\ref{fig:interval_length_varying_pressure_seed129}. Instead of assuming that the pressure is known exactly as was done in Figure~\ref{fig:interval_length_one_fixed_nuance_seed436}, we consider symmetric constraints about the true pressure value, i.e., constraints of the form $x_{21,\mathrm{true}}-\delta \le x_{21} \le x_{21,\mathrm{true}}+\delta$ for various~$\delta$. As expected, we observe that tighter constraints on pressure translate into shorter intervals for~$X_\mathrm{CO2}$. For example, knowing the pressure to within $\pm 3$~hPa lets us decrease the average $X_\mathrm{CO2}$ interval length from 11.19~ppm to 7.55~ppm. Knowing the surface pressure to within such, or even higher, accuracy is not implausible as there are other, complementary observing systems, such as ground-based weather stations, that are capable of providing pressure information within such limits.

We remark that the interval lengths in Figures~\ref{fig:interval_length_one_fixed_nuance_seed436} and \ref{fig:interval_length_varying_pressure_seed129} are averages over 100 noise realizations for the $\bm{x}$ realization corresponding to nominal operational coverage in Table~\ref{tab:coverage_comparison_single_sounding}; see also Figure~S8 in the Supplement \cite{Patil2021Supp}. We have also studied other $\bm{x}$'s from the generative model and found qualitatively similar results.

\subsection{Probabilistic constraints and interval length optimization}

We analyzed above the effect of additional deterministic constraints on pressure. However, such constraints might not always be available with full certainty; instead, we might know that they hold with high probability. This is the case, for example, when a frequentist confidence interval is available from another observing system. In this section, we show how to incorporate such probabilistic constraints within the proposed method while still maintaining finite-sample coverage.

\subsubsection{Coverage calibration} \label{sec:calibration}

To explain the key idea, imagine that instead of having deterministic constraints such as $x_{i,\mathrm{true}}-\delta \le x_{i} \le x_{i,\mathrm{true}}+\delta$, we have confidence intervals for one or more of the $x_i$'s such that $\underline{x_i}(\alpha_i) \le x_i \le \overline{x_i}(\alpha_i)$ with frequentist coverage at least $(1 - \alpha_i)$, i.e., $\prob(x_i \in \left[\underline{x_i}(\alpha_i), \overline{x_i}(\alpha_i)\right]) \geq 1-\alpha_i$. We can then construct a $(1 - \alpha)$ confidence interval for the quantity of interest $\theta$ by running the proposed retrieval procedure with these probabilistic constraints at an internal confidence level $(1-\gamma)$ chosen so that, accounting for the $\alpha_i$'s, we can still maintain the required nominal coverage. As shown in Appendix~\ref{app:calibProbConstraints}, we can bound the miscoverage probability for the quantity of interest as follows:
\begin{align}
	\prob( \theta \notin \left[ \underline{\theta}, \overline{\theta} \right] )	\le \gamma + \sum\nolimits_{i} \alpha_i, \label{eq:errProbBound}
\end{align}
where $i$ ranges over those variables that have probabilistic constraints. Thus, if we choose $\gamma$ and the $\alpha_i$'s in such a way that $\gamma + \sum_{i} \alpha_i = \alpha$, then we can keep the desired $(1-\alpha)$ coverage for the interval $\left[ \underline{\theta}, \overline{\theta} \right]$.

\subsubsection{Demonstration with pressure intervals}

\begin{figure}[t]
	\centering
	\subfigure[Joint effect of pressure and internal confidence level]{
		\includegraphics[height=0.35\columnwidth, clip=true, trim = 0 0 0 0.87cm]{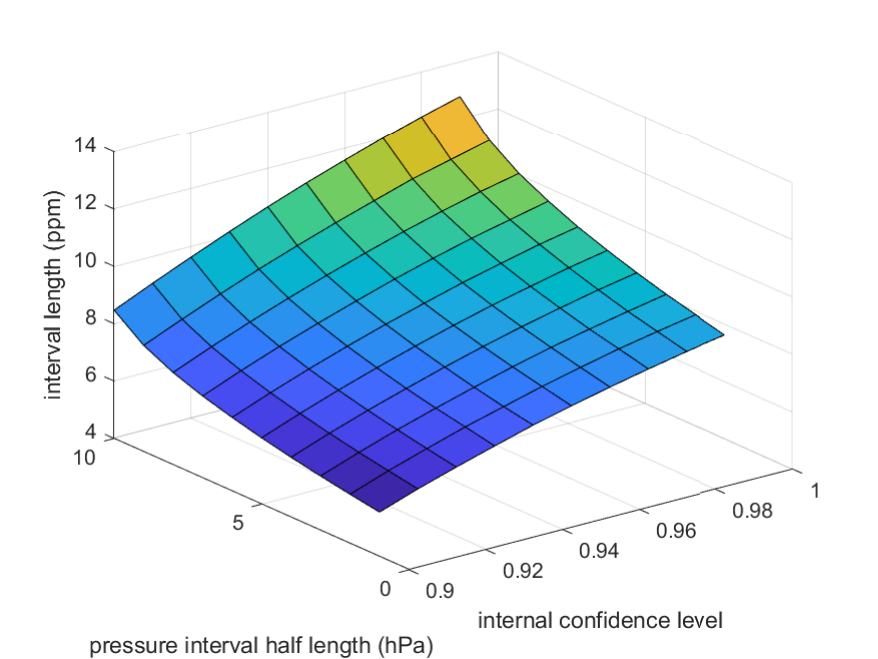}
		\label{fig:interval_length_varying_pressure_coverage_seed255}}
	\subfigure[Interval length optimization]{
		\includegraphics[height=0.35\columnwidth]{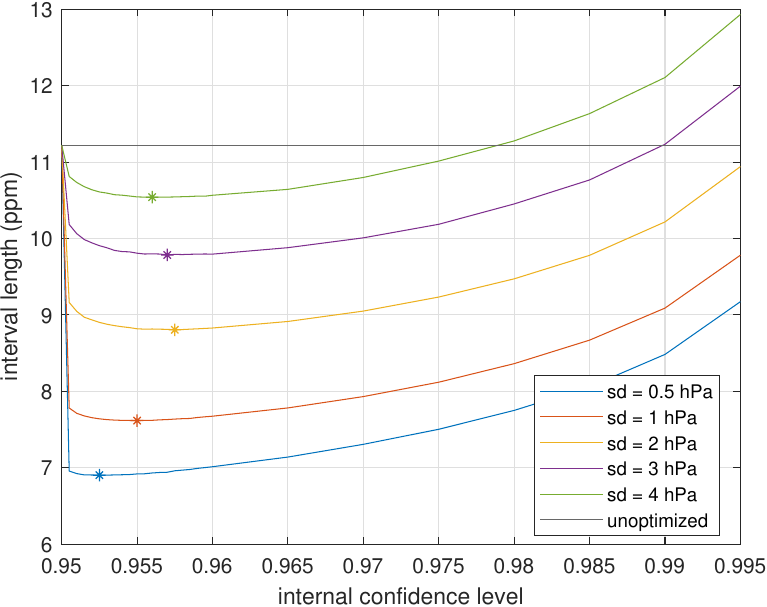}
		\label{fig:interval_length_coverage_slices_seed255}}
	\caption{Figure~\subref{fig:interval_length_varying_pressure_coverage_seed255} shows the joint effect of pressure interval length and internal confidence level on the average $X_\mathrm{CO2}$ interval length. Figure~\subref{fig:interval_length_coverage_slices_seed255} shows the optimization of the average $X_\mathrm{CO2}$ interval length for various pressure standard errors by trading off the internal confidence level $(1-\gamma)$ for the confidence level of the pressure interval $(1-\alpha_{21})$. The values that optimize the interval length are marked with asterisks.}
	\label{fig:interval_length_optimization_pressure}
\end{figure}

Since deterministic constraints on the pressure variable $x_{21}$ provided a gain in the interval length, we now analyze the effect of probabilistic constraints on that variable. This demonstrates the simplest application of Equation~\eqref{eq:errProbBound} in a case where there is a probabilistic constraint on a single variable only. We therefore need to choose $\gamma$ and $\alpha_{21}$ such that $\gamma + \alpha_{21} = \alpha$, where we set $\alpha = 0.05$ to obtain a 95\% final interval for $X_\mathrm{CO2}$. By Equation~\eqref{eq:errProbBound}, any positive $\gamma$ and $\alpha_{21}$ summing to 0.05 will  give a valid final interval, but an optimal choice is such that it minimizes the final interval length. To start investigating the dependence of the $X_\mathrm{CO2}$ interval length on these choices, Figure~\ref{fig:interval_length_varying_pressure_coverage_seed255} shows how the length the pressure interval and the confidence level $(1-\gamma)$ of the $X_\mathrm{CO2}$ interval jointly affect the average $X_\mathrm{CO2}$ interval length. Using Figure~\ref{fig:interval_length_varying_pressure_coverage_seed255}, we can set the internal confidence level $(1-\gamma)$ to a value larger than $95\%$ to account for the coverage probability $(1-\alpha_{21})$ of the pressure interval. To optimize this choice, we need to relate $\alpha_{21}$ to the length of the pressure interval. In this study, we assume that there is a pressure sensor that provides pressure observations $\hat{x}_{21}$ following the Gaussian distribution $\mathcal{N}(x_{21},\sigma_{21}^2)$. We then assume that the pressure intervals are $(1-\alpha_{21})$ standard error intervals of length $2 z_{1-\alpha_{21}/2} \sigma_{21}$, where $\sigma_{21}$ is the pressure standard error.

For a given $\sigma_{21}$, we can then trace through Figure~\ref{fig:interval_length_varying_pressure_coverage_seed255} for various $\gamma$ and the corresponding $\alpha_{21}$ and record the final $X_\mathrm{CO2}$ interval length. Figure \ref{fig:interval_length_coverage_slices_seed255} shows examples of this for pressure standard errors $\sigma_{21}$ ranging from 0.5~hPa to 4~hPa. Each curve represents the average interval length for the proposed method as a function of $\gamma$ and can be used to choose $\gamma$ such that the final interval length is optimized. Along each curve, we have indicated this optimal internal confidence level. We observe that for moderate values of the pressure standard error, the optimal internal confidence level is greater than 95\%, but when the pressure standard error is either very small or very large, the optimal internal confidence level approaches 95\%. This happens because when the pressure standard error is very small, it is almost as good as using the exact pressure value, while when the standard error is very large, it is almost as good as not using any additional constraints on pressure besides the non-negativity constraint.

Since the proposed interval has variable, data-dependent length, there is an important subtlety in that the above interval length optimization must be done without using the observed data $\y$ so as to guarantee the coverage in Equation~\eqref{eq:errProbBound}. In addition, we would ideally like to optimize the average interval length which cannot be done based on a single $\y$. We therefore need a candidate state vector $\x$ that can be used to calculate average interval lengths which are then used as the basis for the length optimization. Since we found in Section~\ref{sec:numerical} that the average interval lengths are not very sensitive to the choice of $\x$, it suffices to have a~reasonable ansatz for $\x$. Luckily, we already have that in the prior mean $\bm{\mu}_a$ of the operational method. In this study, we therefore set the state vector $\bm{x}$ equal to $\bm{\mu}_a$ for the interval length optimization. The interval lengths shown in Figures~\ref{fig:interval_length_varying_pressure_coverage_seed255} and \ref{fig:interval_length_coverage_slices_seed255} were obtained as averages over 100 noise realizations for this choice of $\x$ and for pressure intervals centered at $x_{21} = \mu_{a,21}$.

We now proceed to empirically verify the coverage of the final intervals constructed as described above. The length optimization phase can be run based on $\bm{\mu}_a$ and $\sigma_{21}$ before seeing $\y$. This leads to an optimal choice of $\gamma$ and $\alpha_{21}$ irrespective of $\y$, and, fixing these values, one can then check the coverage and length of the intervals for multiple $\y$ realizations corresponding to a fixed $\x$. We use the $\bm{x}$ that provides nominal coverage for the operational retrieval method in Table~\ref{tab:coverage_comparison_single_sounding} as the state vector for this evaluation. While the length optimization was done without fluctuating the pressure intervals, the coverage study also accounts for the variation of the pressure intervals by simulating intervals of the form $[\hat{x}_{21}-z_{1-\alpha_{21}/2}\sigma_{21},\hat{x}_{21}+z_{1-\alpha_{21}/2}\sigma_{21}]$ for the optimized $\alpha_{21}$ and for $\hat{x}_{21} \sim \mathcal{N}(x_{21},\sigma_{21}^2)$ independently of $\bm{\eps}$. The results are given in Table~\ref{tab:coverage_evaluation_proposed_extension} which shows the optimized confidence levels $(1-\gamma)$ and $(1-\alpha_{21})$ as well as the empirical coverage and average length of the final $X_\mathrm{CO2}$ intervals based on 10\,000 realizations. We observe that the intervals maintain the 95\% coverage guarantee while significantly reducing the final interval length. The amount of gain provided by the pressure information depends on the level of uncertainty in the pressure intervals. In particular, for pressure standard error of 0.5~hPa, we are able to reduce the average interval length to 6.89~ppm from the original 11.19~ppm. The final intervals are somewhat conservative due to the slack in the inequality in Equation~\eqref{eq:errProbBound}. Notice also that the interval lengths predicted by the prior-based optimization in Figure~\ref{fig:interval_length_coverage_slices_seed255} match well with the final values in Table~\ref{tab:coverage_evaluation_proposed_extension} even though this evaluation is for a different $\x$.

\begin{table}[t] \label{tab:coverage_evaluation_proposed_extension}
	\centering
	\caption{\noindent Optimized confidence levels, final empirical coverage and average length for the $X_\mathrm{CO2}$ intervals incorporating probabilistic pressure constraints at various levels of pressure standard error. The internal and pressure confidence levels are chosen so that the final interval has at least 95\% coverage.}
	\begin{tabular}{ c c c c c } 
		\toprule
		pressure & internal & pressure & final & interval \\
		std.\ err.  & conf.\ level & conf.\ level & empirical & length  \\
		(hPa) & $(1-\gamma)$ & $(1-\alpha_{21})$ & coverage & (ppm) \\
		\midrule
		0.5 & 0.9525 & 0.9975  & 0.9741 & 6.89 \\ 
		1 & 0.9550 & 0.9950 & 0.9782 & 7.61 \\ 
		2 & 0.9575 & 0.9925 & 0.9743 & 8.80 \\
		3 & 0.9570 & 0.9930 & 0.9684 & 9.71 \\
		4 & 0.9560 & 0.9940 & 0.9629 & 10.32 \\
		\bottomrule
	\end{tabular}
\end{table}

\section{Conclusions and outlook}
\label{sec:conclusion}

Our focus on the frequentist properties of the uncertainty estimates is one of the main differences between this work and much of the other related work on uncertainty quantification in remote sensing, which tends to predominantly focus on Bayesian construction and evaluation of uncertainties. The frequentist and Bayesian paradigms answer fundamentally different questions about the unknown parameter $\theta$, and as is well known from the extensive discussion in the literature (see \cite{tenorio2017,Biegler2011,stark2011,Scales2001,Evans2002} for references specific to inverse problems), both approaches are valuable in their own right. The question we set out to answer is the following: Given a fixed state of the atmosphere corresponding to a given satellite overpass, what are the repeated sampling properties of the uncertainty intervals, when the repetitions are over the instrument noise~$\bm{\varepsilon}$? Hence, most of our probabilities and expectations are taken with respect to the noise $\bm{\varepsilon}$, while some previous works take expectations over both $\bm{\varepsilon}$ and $\bm{x}$ \cite{hobbs2017,nguyen2019}. In the case of the operational retrieval, our studies constitute a frequentist evaluation of the underlying Bayesian procedure \cite{Bayarri2004}. Arguably, properties calculated with respect to $\bm{\varepsilon}$ are potentially more relevant for downstream scientific use, where, for example, carbon flux estimates use OCO-2 data to gain information about the instantaneous state of the atmosphere corresponding to a particular $\bm{x}$ instead of an average~$\bm{x}$.

It is important to clarify that there is a difference between frequentist and Bayesian criteria for evaluating uncertainties and frequentist and Bayesian constructions of uncertainties. Indeed, some Bayesian constructs can have desirable frequentist properties, while some frequentist constructs may have unexpectedly poor frequentist properties. The results in this paper show that the standard operational retrieval procedure does not fall in the former category, but alternative Bayesian constructs might have improved frequentist properties \cite{Berger2006,Kass1996}. Similarly, standard frequentist approaches to ill-posed inverse problems may have poor frequentist coverage performance \cite{kuusela2015,kuusela2016,kuusela2017}. For example, a variant of penalized maximum likelihood (or equivalently, Tikhonov regularization or ridge regression) would have exactly the same point estimator $\hat{\theta}$ as the operational retrieval but with uncertainty quantified using the standard error interval \eqref{eq:stdErrInt} instead of the credible interval \eqref{eq:XCO2_cred_int}. The resulting interval has always coverage less than $(1-\alpha)$ \cite[Section~6.4.2]{kuusela2016}. In this sense, the operational Bayesian retrieval has better frequentist performance than this alternative frequentist construct (see \cite{Nychka1988,Ruppert2000} for a similar observation in spline smoothing). The difference in the frequentist performance of the two methods considered in this paper is therefore less about the difference between frequentist and Bayesian constructs and more about the difference between explicit and implicit regularization. The proposed method achieves good frequentist calibration because it is implicitly regularized by the functional and the constraints, while the operational retrieval has poor calibration because of the explicit regularization from the prior, and the same conclusion would be true for other explicitly regularized methods. Similarly, it might be possible to obtain an implicitly regularized Bayesian construction by considering a uniform or nearly-uniform vague prior consistent with the available physical constraints.

To interpret the frequentist $X_\mathrm{CO2}$ intervals, it is crucial to understand that the $(1-\alpha)$ coverage property not only holds for a collection of intervals from a given sounding location, but also for a collection of intervals arising from soundings at different locations, since the noise $\bm{\varepsilon}$ is independent across soundings. Imagine a collection of, say, 10\,000 sounding locations within an OCO-2 orbit, each with a realization of a 95\% frequentist confidence interval for~$X_\mathrm{CO2}$. Then, we know that roughly 9\,500 of these intervals cover their true $X_\mathrm{CO2}$ values, and the coverage/non-coverage pattern should not have any apparent spatial structure. It is foreseeable that such intervals could be used to produce rigorous uncertainties in downstream scientific tasks by, for example, using techniques similar to those described here for~$X_\mathrm{CO2}$. Our grid sounding experiments show that the same conclusion does not necessarily hold for the operational retrievals. Let $I_k$ be the indicator random variable indicating whether the $k$th interval covers its true $X_\mathrm{CO2}$ value, where $k$ ranges over the spatial sounding locations within the orbit. For well-calibrated frequentist intervals, the $I_k$'s are independent and identically distributed across the sounding locations, while in the case of the operational retrievals, the $I_k$'s are independent across the sounding locations, but no longer identically distributed. Instead, the coverage probability $\mathbb{P}_{\bm{\eps}}(I_k = 1)$ varies throughout the orbit in a systematic, spatially coherent way, so that in some parts of the orbit perhaps 85\% of the intervals are expected to cover while in other parts maybe 99\% of the intervals cover. Since, in the absence of oracle information, there is no straightforward way of telling which of these situations applies in a given region, it is not immediately clear how to properly use such uncertainties in downstream scientific~tasks.

An important question for future work is to understand what implications these conclusions have on CO$_2$ flux estimates. A key question concerns the spatial length scales at which the biases occur in operational $X_\mathrm{CO2}$ point estimates. Our results indicate that there are spatially correlated biases at least at scales of $8 \times 8$ soundings (roughly $8 \text{ km} \times 16 \text{ km}$), which is likely to have implications for regional carbon flux estimates, for example, over urban areas. This conclusion is further corroborated by OCO-2's target mode observations taken on orbits near TCCON sites to assess the empirical behavior of OCO-2 retrievals for individual overpasses \cite{WunchEtAl2017}. Indeed, the retrieval errors for a single target overpass have been found to exhibit substantial spatial correlation~\cite{zhangtechnom}. However, as of now, we do not know whether these bias patterns persist at the scale of a single pixel in global flux inversion models, where the grid resolution is typically of the order of a few hundred kilometers. If they do, then it would be useful to understand how to incorporate our proposed intervals, which do not exhibit spatially correlated offsets, into these models.

An important insight provided by this work is the identification of the surface pressure ($x_{21}$) and the AOD of the first composite aerosol type ($x_{28}$) as key variables affecting the length of the proposed intervals (Figure~\ref{fig:interval_length_one_fixed_nuance_seed436}). This raises the interesting possibility of obtaining more precise $X_\mathrm{CO2}$ estimates by developing Level 2 retrieval methods that combine pressure or aerosol information from other satellites or observing systems with OCO\nobreakdash-2 data. The surface pressure also plays an important role in explaining the performance of the operational retrievals (see Figures~S6--S8 in the Supplement \cite{Patil2021Supp}), which has also been noted in previous studies; see \cite{kiel2019} and the references therein. A more comprehensive analysis of the effects of the different variables, including at different spatial regions, seasons or observing modes, is left as a subject for future work.

In this paper, we have considered a linearized approximation of the nonlinear OCO-2 forward operator. A major topic for future work would be to extend this work to nonlinear forward operators. The basic primal approach from \cite{stark1992}, outlined in Section \ref{sec:proposalPrimal}, still applies in that the extremal values of $\tp{\bm{h}} \bm{x}$ over $\bm{x} \in C \cap D$ would still define valid $(1-\alpha)$ simultaneous confidence intervals. What is not immediately clear, however, is whether the approach from \cite{rust1972, rust1994} for turning these into one-at-a-time intervals still applies. Another major challenge concerns the computation of the intervals since now $D$ can no longer be described by a quadratic inequality and might even be non-convex, depending on the properties of the forward operator. Constructing and characterizing the dual problems would also be substantially more difficult. Nevertheless, since here $\bm{x}$ has a moderate dimension, it is plausible that methods can be developed for solving the primal optimization problems within reasonable time constraints. One potential approach would be to successively linearize the nonlinear part of the programs within an iterative quadratic programming algorithm.

We have shown empirically that the proposed intervals consistently have frequentist coverage very close to the nominal value. In future work, we hope to be able to show what conditions are needed to rigorously guarantee this. As has been pointed out in~\cite{tenorio2007}, the previous proof in \cite{rust1994} appears to be incorrect. The authors in \cite{tenorio2007} even provide a counterexample showing that the intervals can undercover for $\bm{h}$ containing both positive and negative elements. This leaves open the question of whether it is possible to guarantee the coverage when all elements of $\bm{h}$ have the same sign, as is the case here with the $X_\mathrm{CO2}$ functional. If it turns out to be difficult to provide such guarantees, it might be possible to consider alternative definitions of the slack factor $s^2$ so that coverage and other theoretical properties can be proved more~easily.

While they have much better frequentist calibration, the proposed intervals are almost three times as long as the current operational intervals, when only trivial constraints are applied. Therefore, an important question for future work would be to understand what can be said about the optimality of the length of these intervals within the class of methods that provide frequentist coverage guarantees. Donoho introduced in \cite{donoho1994} intervals that are up to a multiplicative factor minimax optimal for this problem among the class of fixed-length intervals with guaranteed coverage. The intervals studied here are variable-length and may hence be shorter than those of \cite{donoho1994}. To the best of our knowledge, minimax optimality of variable-length intervals for this setting is an open problem. Furthermore, instead of minimax, a more appropriate notion of optimality here might be one with respect to a reasonable distribution on $\bm{x}$, such as the operational prior distribution.

\appendix

\section*{Appendices}

Appendices~\ref{app:comp_simplification}--\ref{app:reduction} assume that the forward model has been transformed to have identity covariance, i.e., $\bm{y} \sim \mathcal{N}(\bm{K}\bm{x},\bm{I})$, as described in Section~\ref{sec:proposalOutline}.

\section{Computational simplification} \label{app:comp_simplification}
Consider again the original problem in the primal form to obtain the lower endpoint $\underline{\theta}$:
\begin{equation}
\begin{array}{ll}
\mbox{minimize}   & \h^T \x \\
\mbox{subject to} & \| \y  - \K \x \|^2 \le z_{1-\alpha/2}^2 + s^2, \\
& \A \x \le \bm{b},
\end{array} \label{eq:original}
\end{equation}
where $s^2 = \min_{\x \,:\, \A \x \le \bm{b}} \| \y - \K \x \|^2$.

Let us consider the singular value decomposition $\K = \U \D  \V^T$. We first note that $\| \y - \K \x \|^2 = \| \U^T \y - \D \V^T \x \|^2$, since $\U^T \U = \U \U^T = \I$. Further, as $n > p$, let us denote by $\tilde{\y}_{1:p}$ the first $p$ entries of $\tilde{\y} = \U^T \y$ and by $\tilde{\y}_{p+1:n} $ the rest of the entries of $\tilde{\y}$. Then, we can write $\| \U^T \y - \D \V^T \x \|^2 = \| \tilde{\y}_{1:p} - \D_{1:p,:} \V^T \x \|^2 + \| \tilde{\y}_{p+1:n} \|^2$, where $\D_{1:p,:}$ denotes the first $p$ rows of~$\D$. This suggests a simplification of the primal problem where, instead of \eqref{eq:original}, we solve the following equivalent problem to obtain the lower endpoint $\underline{\theta}$:
\begin{equation}
\begin{array}{ll}
\mbox{minimize}   & \h^T \x \\
\mbox{subject to} &  \| \tilde{\y}_{1:p} - \D_{1:p,:} \V^T \x \|^2 \le z_{1-\alpha/2}^2 + \tilde{s}^2, \\
& \A \x \le \bm{b},
\end{array} \label{eq:simplified}
\end{equation}
where now $\tilde{s}^2 = \min_{\x \,:\, \A \x \le \bm{b}} \| \tilde{\y}_{1:p} - \D_{1:p,:} \V^T \x \|^2$. This is equivalent to the original problem because $s^2 = \tilde{s}^2 + \| \tilde{\y}_{p+1:n} \|^2$. When $n \gg p$, solving problem \eqref{eq:simplified}, including the associated slack problem, is much faster than solving problem \eqref{eq:original}, because the norms involve $p$-variate vectors instead of $n$-variate vectors. An analogous simplification can obviously be used with the upper endpoint $\overline{\theta}$ as well. These simplifications proved crucial for our ability to perform the empirical coverage studies presented in this paper.

\section{Dual derivation} \label{app:dual_derivation}
Consider the primal optimization problem to obtain the lower endpoint $\underline{\theta}$:
\begin{equation}
\begin{array}{ll}
\mbox{minimize}   & \h^T \x \\
\mbox{subject to} & \| \y  - \K \x \|^2 \le z_{1 - \alpha/2}^2 + s^2, \\
& \A \x \le \bm{b},
\end{array}
\end{equation}
where $s^2$ is the slack factor. For notational convenience, we let $z_{1 - \alpha/2}^2 + s^2 = q^2$.

We first write an equivalent problem as follows:
\begin{equation}
\begin{array}{ll}
\mbox{minimize}   & \h^T \x \\
\mbox{subject to} & \y  - \K \x = \bm{r}, \\
& \| \bm{r} \|^2 \le q^2, \\
& \A \x \le \bm{b},
\end{array}
\end{equation}
where the optimization is now over both $\x$ and $\bm{r}$.

The Lagrangian of the above problem can be written as
\begin{equation}
	L(\x, \bm{r}, \bm{w}, \lambda, \bm{c}) = 
	\h^T \x + \bm{w}^T (\y - \K \x - \bm{r}) + \lambda (\| \bm{r}\|^2 - q^2) + \bm{c}^T (\bm{A} \x - \bm{b}),
\end{equation}
where $\bm{w}$, $\lambda \ge 0$ and $\bm{c} \ge \bm{0}$ are dual variables \cite{boyd2004}.

The dual function is obtained by minimizing the Lagrangian with respect to the primal variables $\x$ and $\bm{r}$:
\begin{equation}
	g(\bm{w}, \lambda, \bm{c}) = \inf_{\x, \bm{r}} L(\x, \bm{r}, \bm{w}, \lambda, \bm{c}).
\end{equation}
We first rewrite the Lagrangian to group the terms corresponding to $\x$ and $\bm{r}$ together:
\begin{equation}
L(\x, \bm{r}, \bm{w}, \lambda, \bm{c}) = 
(\h - \K^T \bm{w} + \bm{A}^T \bm{c})^T \x - \bm{w}^T \bm{r} + \lambda \| \bm{r}\|^2 + \bm{w}^T \y - \lambda q^2 - \bm{c}^T \bm{b}.
\end{equation}
Next, we note that we can restrict ourselves to the case where $\h - \K^T \bm{w} + \bm{A}^T \bm{c} = \bm{0}$, since otherwise the Lagrangian is unbounded below as a linear function in $\x$. By minimizing with respect to $\bm{r}$ and substituting back, we obtain the dual function
\begin{equation}
g(\bm{w}, \lambda, \bm{c}) = - \frac{1}{4\lambda} \| \bm{w} \|^2 + \bm{w}^T \y - \lambda q^2 - \bm{c}^T \bm{b},
\end{equation}
where $\h - \K^T \bm{w} + \bm{A}^T \bm{c} = \bm{0}$, $\lambda \geq 0$ and $\bm{c} \geq \bm{0}$. The dual optimization problem is then the problem of maximizing $g(\bm{w}, \lambda, \bm{c})$ with respect to the dual variables $\w$, $\lambda$ and $\bm{c}$. Maximization with respect to $\lambda$ can be carried out in closed form. We can thus eliminate $\lambda$ to obtain the following dual problem for the remaining variables $\w$ and $\bm{c}$:
\begin{equation}
\begin{array}{ll}
\mbox{maximize}   & \w^T \y - \sqrt{z_{1 - \alpha/2}^2 + s^2} \| \w \| - \bm{b}^T \bm{c}\\
\mbox{subject to} & \h + \A^T \bm{c} - \K^T \w = \bm{0}, \\
& \bm{c} \ge \bm{0},
\end{array}
\end{equation}
which gives us Equation~\eqref{eq:dualLb}. The dual problem \eqref{eq:dualUb} corresponding to the upper endpoint $\overline{\theta}$ follows from an analogous derivation.

\section{Coverage for fixed dual variables} \label{sec:dualCoverage}
Assume $\bm{y} \sim \mathcal{N}(\bm{K}\bm{x},\bm{I})$ with functional of interest $\theta = \bm{h}^T \bm{x}$ and state vector $\bm{x}$ satisfying $\bm{A}\bm{x} \leq \bm{b}$. Consider a lower endpoint of the form $\underline{\theta} = \w^T \y - z_{1- \alpha/2} \| \w \| - \bm{b}^T \bm{c}$ for some fixed $\w$ and $\bm{c}$ satisfying the constraints in program~\eqref{eq:dualLb}. We can bound the miscoverage probability as follows:
{\allowdisplaybreaks
\begin{align*}
\prob_{\bm{\eps}}(\underline{\theta} \ge \theta) &= \prob_{\bm{\eps}}( \w^T \y - z_{1 - \alpha/2} \|\w\| - \bm{b}^T \bm{c} \ge \theta ) \\
& = \prob_{\bm{\eps}}( \w^T \y - \w^T \K \x -  z_{1 - \alpha/2} \|\w\| \ge \h^T \x - \w^T \K \x + \bm{b}^T \bm{c} ) \\
& \overset{(1)}{\le} \prob_{\bm{\eps}}( \w^T \y - \w^T \K \x -  z_{1 - \alpha/2} \|\w\| \ge  \h^T \x - \w^T \K \x + \bm{c}^T \A \x ) \\
& \overset{(2)}{=} \prob_{\bm{\eps}}( \w^T \y - \w^T \K \x -  z_{1 - \alpha/2} \|\w\| \ge 0 ) \overset{(3)}= \alpha/2,
\end{align*}}where $(1)$ follows from the constraints $\A \x \le \bm{b}$ and $\bm{c} \ge \0$; $(2)$ uses the fact that $\w$ and $\bm{c}$ need to satisfy the constraint $\h + \A^T \bm{c} - \K^T \w = \bm{0}$; and $(3)$ follows from $\w^T \y \sim \N(\w^T \K \x, \|\w\|^2)$ for any fixed $\w$. Thus, for fixed $\w$ and $\bm{c}$ that satisfy the constraints, we have $\prob_{\bm{\eps}}(\underline{\theta} \ge \theta) \le \alpha/2$.

\section{Simplification with full column rank and no constraints} \label{app:reduction}
We prove in this section that when $\mathrm{rank}(\bm{K}) = p$ and there are no external constraints on $\bm{x}$, the solutions of problems \eqref{eq:primalMin} and \eqref{eq:primalMax} yield the interval $[\hat{\theta}_\mathrm{LS} - z_{1 - \alpha/2} \,\mathrm{se}(\hat{\theta}_\mathrm{LS}), \hat{\theta}_\mathrm{LS} + z_{1 - \alpha/2} \,\mathrm{se}(\hat{\theta}_\mathrm{LS})]$, where $\hat{\theta}_\mathrm{LS} = \h^T \hat{\x}_\mathrm{LS}$ is the plug-in estimator of $\theta$, $\hat{\x}_\mathrm{LS} = (\K^T \K)^{-1} \K^T \y$ is the unregularized least-squares estimator of $\x$ and $\mathrm{se}(\hat{\theta}_\mathrm{LS}) = \sqrt{\h^T (\K^T\K)^{-1} \h}$ is the standard error of $\hat{\theta}_\mathrm{LS}$.

Consider the lower endpoint of the interval. In the absence of external constraints on $\bm{x}$, the optimization problem \eqref{eq:primalMin} reduces to
\begin{equation} \label{eq:primalMin_unconstrained}
\begin{array}{ll}
\mbox{minimize}   & \h^T \x \\
\mbox{subject to} & \| \y  - \K \x \|^2 \le z_{1- \alpha/2}^2 + s^2,
\end{array}
\end{equation}
where the slack factor $s^2$ is now defined as the objective function value of the corresponding unconstrained least-squares problem:
\begin{equation}
s^2 = \underset{\bm{x}}{\mbox{min}}\ \| \y  - \K \x \|^2.
\end{equation}
Since we assume that $\bm{K}$ has full column rank, the solution to the above problem is exactly $\hat{\x}_\mathrm{LS} = (\K^T \K)^{-1} \K^T \y$. Plugging in this value of $\hat{\x}_\mathrm{LS}$, we obtain that the squared slack is given by the residual sum of squares
\begin{equation}
	s^2 = \| \y  - \K  (\K^T \K)^{-1} \K^T \y\|^2.
\end{equation}
We can then write the constraint in problem~\eqref{eq:primalMin_unconstrained} as follows:
\begin{equation}
	\| \y  - \K \x \|^2 \le z_{1- \alpha/2}^2 + \| \y  - \K  (\K^T \K)^{-1} \K^T \y\|^2.
\end{equation}
We can further manipulate the difference
\begin{align}
\| \y  - \K \x \|^2 -  \| \y  - \K  (\K^T \K)^{-1} \K^T \y\|^2
&= \x^T \K^T \K \x - 2 \y^T \K \x + \y^T \K (\K^T \K)^{-1} \K^T \y \notag \\
&= \| \x - \hat{\x}_\mathrm{LS} \|_{\K^T \K}^2,
\end{align}
where we have used the weighted-norm notation $\| \x \|_{\bm{A}} = \sqrt{\x^T \bm{A} \x}$, to arrive at the following program for the lower endpoint of the interval:
\begin{equation}
\begin{array}{ll}
\mbox{minimize}   & \h^T \x \\
\mbox{subject to} & \| \x - \hat{\x}_\mathrm{LS} \|_{\K^T \K}^2 \le z_{1- \alpha/2}^2.
\end{array}
\end{equation}
We proceed to show that the optimal value of this problem is given by $\hat{\theta}_\mathrm{LS} - z_{1 - \alpha/2}\mathrm{se} \, (\hat{\theta}_\mathrm{LS})$. We begin by writing down the Karush--Kuhn--Tucker (KKT) conditions \cite{boyd2004} of the problem. The Lagrangian of the problem is given by
\begin{equation}
	L(\x, \lambda) = \h^T \x + \lambda \left( \| \x - \hat{\x}_\mathrm{LS} \|_{\K^T \K}^2 - z_{1- \alpha/2}^2 \right),
\end{equation}
where $\lambda \geq 0$ is a dual variable. The KKT conditions for the primal and dual optimal pair $(\x^\star, \lambda^\star)$ are thus
\begin{equation}
\h = - 2 \lambda^\star \K^T \K (\x^\star - \hat{\x}_\mathrm{LS}),
\end{equation}
using first-order optimality with respect to $\x$, along with
\begin{equation}
	\lambda^\star \left( \| \x^\star - \hat{\x}_\mathrm{LS} \|_{\K^T \K}^2 - z_{1- \alpha/2}^2 \right) = 0
\end{equation}
from the complementary slackness condition. We find that
$\lambda^\star = \frac{1}{2 z_{1-\alpha/2}} \| \h \|_{(\K^T\K)^{-1}} \ge 0$
and $\x^\star = \hat{\x}_\mathrm{LS} - \frac{z_{1- \alpha/2}}{\| \h \|_{(\K^T\K)^{-1}}} (\K^T \K)^{-1} \h$ satisfy the KKT conditions and therefore provide a primal-dual optimal pair. Substituting the value of $\x^\star$ into the objective, we arrive at the desired lower endpoint. The upper endpoint results from a similar argument.

\section{Coverage of Gaussian central credible intervals} \label{app:credibCoverage}
This section provides derivation of the frequentist coverage of the credible interval \eqref{eq:XCO2_cred_int} used in the operational retrievals. Since $\hat{\theta}$ is an affine transformation of $\bm{y}$, it is Gaussian with mean $\mathbb{E}_{\bm{\eps}}[\hat{\theta}]$ and variance $\mathrm{var}_{\bm{\varepsilon}}(\hat{\theta})$. Therefore, $\big(\hat{\theta}-\theta-\mathrm{bias}(\hat{\theta})\big)/\mathrm{se}(\hat{\theta})$ has standard Gaussian distribution. Then,
\begin{align*}
\mathbb{P}_{\bm{\eps}}(\theta \in [\underline{\theta}, \overline{\theta}]) &= \mathbb{P}_{\bm{\eps}}( \hat{\theta} - z_{1-\alpha/2} \sigma \leq \theta \leq \hat{\theta} + z_{1-\alpha/2} \sigma ) \\
&= \mathbb{P}_{\bm{\eps}}( - z_{1-\alpha/2} \sigma \leq \hat{\theta} - \theta \leq z_{1-\alpha/2} \sigma ) \\
&= \mathbb{P}_{\bm{\eps}}\left( -\frac{\mathrm{bias}(\hat{\theta})}{\mathrm{se}(\hat{\theta})} - z_{1-\alpha/2} \frac{\sigma}{\mathrm{se}(\hat{\theta})} \leq \frac{\hat{\theta} - \theta - \mathrm{bias}(\hat{\theta})}{\mathrm{se}(\hat{\theta})} \leq -\frac{\mathrm{bias}(\hat{\theta})}{\mathrm{se}(\hat{\theta})} + z_{1-\alpha/2} \frac{\sigma}{\mathrm{se}(\hat{\theta})} \right) \\
&= \Phi\left( -\frac{\mathrm{bias}(\hat{\theta})}{\mathrm{se}(\hat{\theta})} + z_{1-\alpha/2} \frac{\sigma}{\mathrm{se}(\hat{\theta})} \right) - \Phi\left( -\frac{\mathrm{bias}(\hat{\theta})}{\mathrm{se}(\hat{\theta})} - z_{1-\alpha/2} \frac{\sigma}{\mathrm{se}(\hat{\theta})} \right) \\
&= \Phi\left( \frac{\mathrm{bias}(\hat{\theta})}{\mathrm{se}(\hat{\theta})} + z_{1-\alpha/2} \frac{\sigma}{\mathrm{se}(\hat{\theta})} \right) - \Phi\left( \frac{\mathrm{bias}(\hat{\theta})}{\mathrm{se}(\hat{\theta})} - z_{1-\alpha/2} \frac{\sigma}{\mathrm{se}(\hat{\theta})} \right),
\end{align*}
using $\Phi(x) = 1 - \Phi(-x)$ to obtain the last equality. This establishes Equation~\eqref{eq:operational_coverage}.

\section{Miscoverage probability with probabilistic constraints} \label{app:calibProbConstraints} Given the setup in Section~\ref{sec:calibration}, we can bound the error probabilities as follows:
	{\allowdisplaybreaks
	\begin{align}
	\prob( \theta \notin \left[ \underline{\theta}, \overline{\theta} \right] )
	&= \prob( \theta \notin \left[ \underline{\theta}, \overline{\theta} \right], x_{i} \in \left[\underline{x_i}(\alpha_i), \overline{x_i}(\alpha_i) \right] \text{for all } i) \notag \\ &\qquad\qquad + \prob( \theta \notin \left[ \underline{\theta}, \overline{\theta} \right], x_{i} \notin \left[\underline{x_i}(\alpha_i), \overline{x_i}(\alpha_i) \right] \text{for some } i) \notag \\
	&= \prob( \theta \notin \left[ \underline{\theta}, \overline{\theta} \right] ~|~ x_{i} \in \left[\underline{x_i}(\alpha_i), \overline{x_i}(\alpha_i) \right] \text{for all } i) \cdot \prob(x_{i} \in \left[\underline{x_i}(\alpha_i), \overline{x_i}(\alpha_i) \right] \text{for all } i) \notag \\ &\qquad\qquad + \prob( \theta \notin \left[ \underline{\theta}, \overline{\theta} \right], x_{i} \notin \left[\underline{x_i}(\alpha_i), \overline{x_i}(\alpha_i) \right] \text{for some } i) \notag \\
	&\le \prob( \theta \notin \left[ \underline{\theta}, \overline{\theta} \right] ~|~ x_{i} \in \left[\underline{x_i}(\alpha_i), \overline{x_i}(\alpha_i) \right] \text{for all } i) \notag \\&\qquad\qquad + \prob( x_{i} \notin \left[\underline{x_i}(\alpha_i), \overline{x_i}(\alpha_i) \right] \text{for some } i ) \notag \\
	&\le \prob( \theta \notin \left[ \underline{\theta}, \overline{\theta} \right] ~|~ x_{i} \in \left[\underline{x_i}(\alpha_i), \overline{x_i}(\alpha_i) \right] \text{for all } i) \notag \\&\qquad\qquad + \sum\nolimits_{i} \prob( x_{i} \notin \left[\underline{x_i}(\alpha_i), \overline{x_i}(\alpha_i) \right] ) \notag \\
	&\le \gamma + \sum\nolimits_{i} \alpha_i, \label{eq:calibProbConstraints}
	\end{align}}where $i$ ranges over those variables that have probabilistic constraints.

We note that when this framework is used to incorporate probabilistic constraints on multiple variables, using the union bound to control the miscoverage probability, as is done in Equation~\eqref{eq:calibProbConstraints}, might be loose and additional structure among the probabilistic constraints, such as independence, could provide additional gain.

\section*{Acknowledgments} We are grateful to Amy Braverman, Mike Gunson, Ryan Tibshirani, Sivaraman Balakrishnan, as well as the participants of the OCO-2 March 2019 Uncertainty Quantification Breakout Meeting, the Statistical and Applied Mathematical Sciences Institute Spatial~X Working Group, and the Carnegie Mellon University Statistical Methods for the Physical Sciences Research Group for helpful discussions and feedback at various stages throughout this work. We would also like to thank the Associate Editor and the two anonymous referees for their insightful feedback that helped us improve various aspects of this~work.

\bibliographystyle{siamplain}
\bibliography{references}



\section*{Supplementary material}

# Supplement to "Objective frequentist uncertainty quantification for atmospheric $CO_2$ retrievals"

Pratik Patil[1,2], Mikael Kuusela[1] and Jonathan Hobbs[3]

[1]Department of Statistics and Data Science, Carnegie Mellon University
[2]Machine Learning Department, Carnegie Mellon University
[3]Jet Propulsion Laboratory, California Institute of Technology

## S1 Supplement to Section 2.1

Figure S1 shows an example sounding for OCO-2.

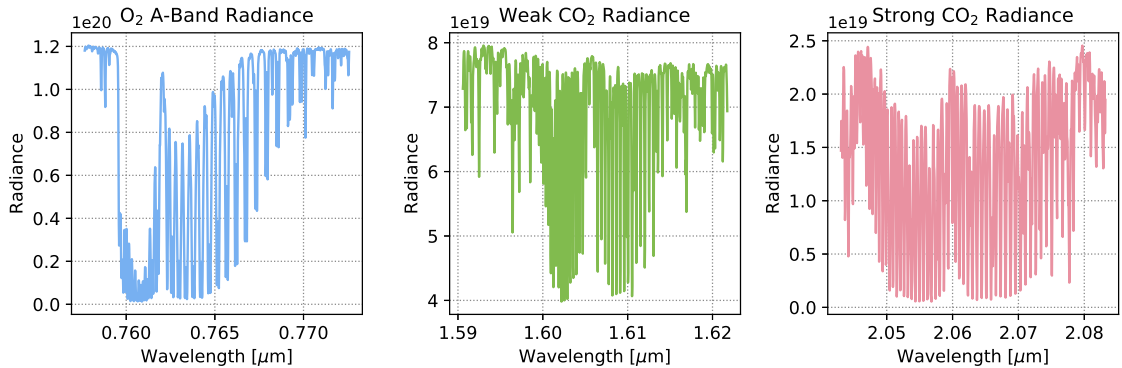

**Figure S1:** Example sounding from OCO-2. A sounding includes 1016 radiances in each of the three infrared spectral bands.

## S2 Supplement to Section 5.1

### S2.1 Description of state vector elements

The specific elements of the state vector $\boldsymbol{x} \in \mathbb{R}^{39}$ are:

- Variables $x_1, \ldots, x_{20}$ are the dry-air mole fractions of atmospheric $CO_2$, i.e., the number of moles of $CO_2$ per one mole of dry air, in parts per million (ppm) at 20 fixed pressure levels. In the sequel, we denote these as levels 1 to 20, with level 1 being highest in the atmosphere (pressure $\sim 0.1$ hPa) and level 20 being the surface.

- Variable $x_{21}$ is the surface air pressure in hPa. It corresponds to the total weight of the air molecules in the atmospheric column.

- Variables $x_{22}, \ldots, x_{27}$ relate to surface albedo, i.e., the fraction of total incoming solar radiation reflected off the Earth's surface. Albedo varies across the three OCO-2 spectral

1

bands and also within each band. In the surrogate model, albedo is modeled as a linear function within each spectral band; see Section B.2 in [1]. The albedo for each band is therefore parameterized by an intercept and a slope which enter the state vector as nuisance variables ($x_{22}$, $x_{24}$, and $x_{26}$ are the three intercepts and $x_{23}$, $x_{25}$, and $x_{27}$ are the slopes).

- Variables $x_{28}, \ldots, x_{39}$ parameterize the atmospheric aerosol concentrations and distributions. The surrogate model assumes that there are 4 aerosol types, which have distinct absorption and scattering properties. The first two are location-dependent composite species. For our investigation, these are sulfate and dust. The latter two are two cloud species, one for ice clouds and another for liquid water clouds [2, 1]. Each type is parameterized by 3 parameters corresponding to the aerosol optical depth (AOD, i.e., the opaqueness of the aerosol species measured as the natural logarithm of the ratio of incoming to transmitted radiation) as well as the altitude and thickness of each aerosol layer. Variables $x_{28}$, $x_{31}$, $x_{34}$, and $x_{37}$ are the log-AOD values, variables $x_{29}$, $x_{32}$, $x_{35}$, and $x_{38}$ are the altitudes, and variables $x_{30}$, $x_{33}$, $x_{36}$, and $x_{39}$ are the log-thicknesses.

### S2.2 Forward model singular values

Figure S2 shows the singular values of the linearized forward model $\boldsymbol{K}$.

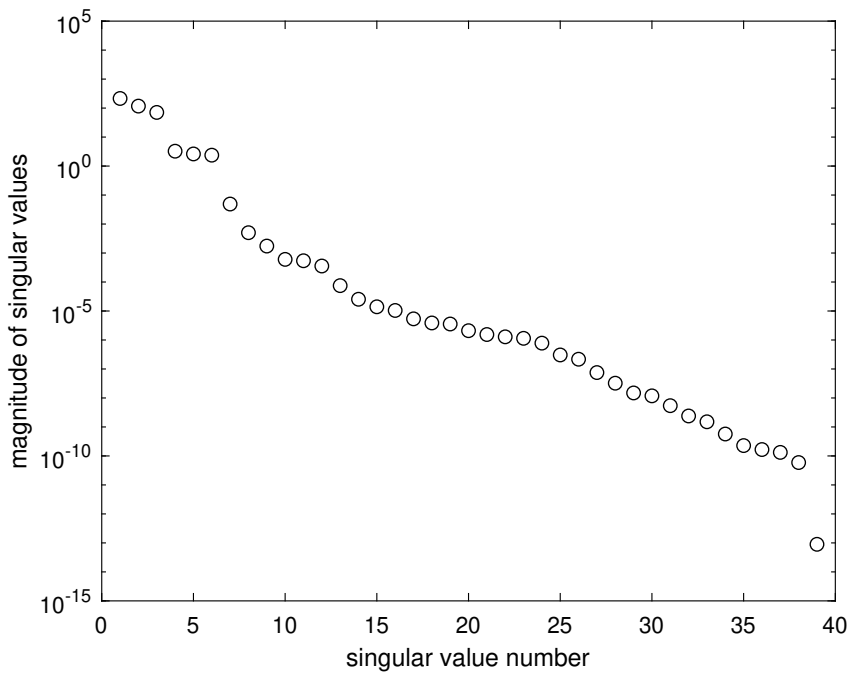

**Figure S2:** Singular value decay of the linearized forward model $\boldsymbol{K}$ near Lamont, OK, in October 2015.

### S2.3 Exploratory analysis of the prior and generative models

In this section, we visualize and describe various components of the generative model and compare those to the prior model. We begin by comparing the prior mean and standard deviation to the true generative model mean and standard deviation for the $CO_2$ part of the state vector

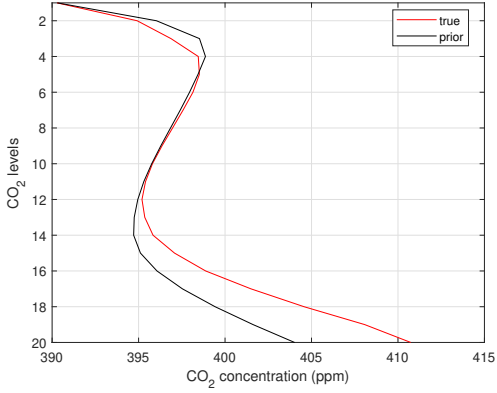

(a) Mean misspecification

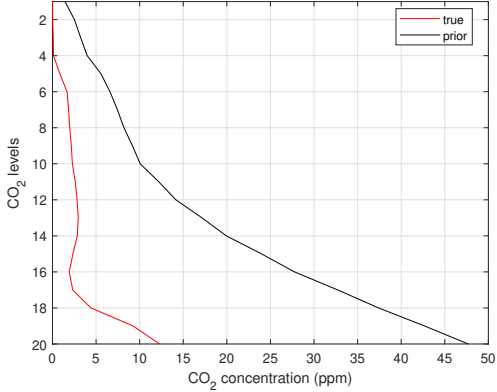

(b) Standard deviation misspecification

**Figure S3:** Visualization of the misspecification of the mean and the standard deviation for the $CO_2$ profile between the true generative process and the prior. Figure (a) visualizes the misspecification in the mean and Figure (b) shows the misspecification in the standard deviation.

**Table S1:** Comparison of the true and prior mean and standard deviation for the nuisance variables $x_i, i = 21, \ldots, 39$. A detailed description of these variables is given in Section S2.1.

| $i$ | true mean (sd) | prior mean (sd) |
|---|---|---|
| | surface pressure | |
| 21 | 972.3235 (1.7508) | 970.6240 (4.000) |
| | albedo | |
| 22 | 0.1267 (0.0204) | 0.1267 (1.0000) |
| 23 | 0.0001 (0.0001) | 0.0000 (0.0005) |
| 24 | 0.2484 (0.0044) | 0.2484 (1.0000) |
| 25 | -0.0001 (0.0000) | 0.0000 (0.0005) |
| 26 | 0.2027 (0.0026) | 0.2027 (1.0000) |
| 27 | -0.0000 (0.0000) | 0.0000 (0.0005) |
| | aerosols | |
| 28 | -3.8786 (0.3380) | -3.7643 (2.0000) |
| 29 | 0.8187 (0.0683) | 0.9000 (0.2000) |
| 30 | -2.4492 (0.0409) | -2.9957 (0.1823) |
| 31 | -6.1959 (0.8492) | -4.7370 (2.0000) |
| 32 | 0.3255 (0.0101) | 0.9000 (0.2000) |
| 33 | -3.9219 (0.0201) | -2.9957 (0.1823) |
| 34 | -4.3980 (0.2477) | -4.3820 (1.8000) |
| 35 | -0.0087 (0.0355) | 0.3000 (0.2000) |
| 36 | -3.2080 (0.0112) | -3.2189 (0.2231) |
| 37 | -5.6803 (0.2517) | -4.3820 (1.8000) |
| 38 | 1.0917 (0.0870) | 0.7500 (0.4000) |
| 39 | -2.3052 (0.0004) | -2.3026 (0.0953) |

in Figure S3. We observe that both the mean and the standard deviation have the largest misspecification near the surface. Table S1 contains the same information for the nuisance variables $x_{21}, \ldots, x_{39}$. Generally speaking, the prior standard deviation is by design larger than that of the true process, which provides some protection against the prior mean misspecification.

We next visualize the correlation structure in the generative process and the prior. Figure S4(a) shows a heat map of the correlation matrix for the true generative process. We observe that the state vector consists of four independent subgroups of variables corresponding to the $CO_2$ profile, surface pressure, albedo variables and aerosol variables. While variables across these groups are uncorrelated, there are large within-group correlations. Figure S4(b) shows a heat map of the correlation matrix for the prior process. Again, the $CO_2$ variables are independent of the nuisance variables, but in this case, there are no correlations between the nuisance variables. Notice also the differences in the correlation structure within the $CO_2$ profile between the generative process and the prior.

Lastly, we visualize the spatial correlation structure for the generative process over an $8 \times 8$

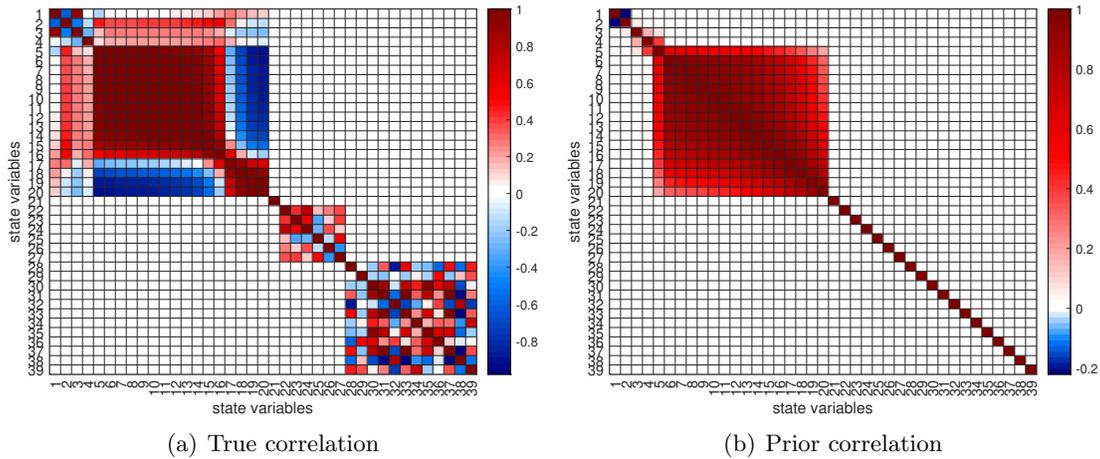

(a) True correlation  (b) Prior correlation

**Figure S4:** Visualization of the misspecification of the correlation structure between the true generative process and the prior. Figure (a) displays the correlation between the state vector elements in the true generative process. Figure (b) shows the same for the prior process.

spatial grid near Lamont, OK. These 64 sounding locations are shown in Figure S5(a) and Figure S5(b) displays the correlations across the locations along the diagonal in Figure S5(a). Nearby state vectors are strongly spatially correlated, but there is a fair amount of decorrelation when moving across the grid.

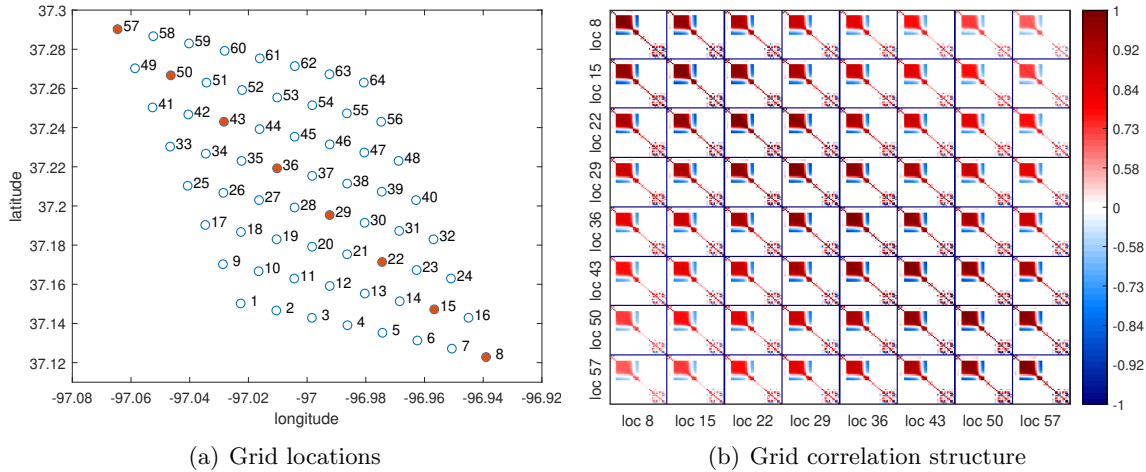

(a) Grid locations  (b) Grid correlation structure

**Figure S5:** Visualization of the correlation structure between the state vector elements for the generative process over an $8 \times 8$ spatial grid. Figure (a) shows the coordinates of the sounding locations near Lamont, OK, with the numbers giving an index for each location. Figure (b) visualizes the correlation structure between the state vector elements across the marked locations along the diagonal in Figure (a). Notice the nonlinear color scale in Figure (b).

## S3 Supplement to Section 5.2: Illustrative instances

We pick some illustrative realizations to visualize the state vectors and confidence intervals produced by the two methods. We choose three representative cases corresponding to the minimum,

nominal and maximum coverage for the operational method in Figure 3(b) which are also rows 1, 8 and 10 in Table 1, respectively.

Figure S6(a) shows the $CO_2$ part of the $\boldsymbol{x}$ instance having the smallest operational coverage. We observe, consistent with Section 5.2.1, that the operational interval is biased upward. It is overoptimistic about the amount of uncertainty, leading it to miss the true $X_{CO2}$ value for this particular $\boldsymbol{\varepsilon}$ realization. From Table 1, we know that this happens for roughly 21% of $\boldsymbol{\varepsilon}$ realizations when the noncoverage probability should ideally be only 5%. The proposed interval, on the other hand, is wider and as a result ends up covering the true $X_{CO2}$ value for the same $\boldsymbol{x}$ and $\boldsymbol{\varepsilon}$ realizations. From Table 1, we know that the noncoverage probability for this interval is 5%, as it should be.

It is quite insightful to investigate the source of the upward bias for this $\boldsymbol{x}$ realization. Contrary to what one might at first imagine, it is not caused by a misspecification of the $CO_2$ profile in the prior. In fact, as shown in Figure S6(a), the prior $CO_2$ profile is very similar to the true $CO_2$ profile in $\boldsymbol{x}$ and the $X_{CO2}$ value implied by the prior is almost the same as the true $X_{CO2}$ value. Instead, it turns out that the bias is primarily caused by an upward fluctuation in the surface pressure variable $x_{21}$, as shown by Figure S6(b). The large positive difference in $x_{21}$ between the true state and the prior gets multiplied by the relatively large positive bias multiplier of this variable (Figure 2(a); see also Section 5.2.1) leading to a large positive contribution to the overall bias. This results in a positively biased operational $X_{CO2}$ point estimate and a miscalibrated confidence interval.

Next, instead of picking an adversarial $\boldsymbol{x}$, Figure S7(a) shows the $\boldsymbol{x}$ realization corresponding to the largest coverage for the operational method. For this $\boldsymbol{x}$, the operational $X_{CO2}$ estimate is effectively unbiased and the operational interval covers the true $X_{CO2}$ value for almost all $\boldsymbol{\varepsilon}$ realizations, resulting in substantial overcoverage. The proposed interval, on the other hand, again has the desired 95% coverage, as indicated by Table 1. For the $\boldsymbol{\varepsilon}$ realization shown in the figure, the operational and proposed intervals both cover the true $X_{CO2}$ value.

Interestingly, the operational $X_{CO2}$ estimator in Figure S7(a) is effectively unbiased even though the prior is badly misspecified for this $CO_2$ profile. Figure S7(b) explains this phenomenon. Due to the small bias multipliers of the low-altitude $CO_2$ values (Figure 2(a)), prior misspecification in the lower half of the $CO_2$ profile creates little bias, while the misspecification in the upper half of the profile is such that it creates both positive and negative contributions to the bias that cancel out. At the same time, there is a consistent positive contribution to the bias from $x_{32}$, which is negated by a downward fluctuation in the pressure variable $x_{21}$ and a consistent negative bias contribution from $x_{33}$.

Finally, we compare in Figure S8(a) the intervals for the $\boldsymbol{x}$ realization that gives nominal coverage for the operational method in Table 1. For this $\boldsymbol{x}$, both intervals cover the true $X_{CO2}$ value for 95% of $\boldsymbol{\varepsilon}$ realizations. The $\boldsymbol{\varepsilon}$ realization shown in the figure has both intervals covering the true $X_{CO2}$ value. The operationally retrieved $CO_2$ profile has fluctuated quite substantially, but the anticorrelated fluctuations cancel out to produce a well-behaved confidence interval for $X_{CO2}$. Overall, the operational method is slightly positively biased for $X_{CO2}$, as expected based on the discussion in Section 5.2.2. Figure S8(b) reveals that the source of this bias is similar to the situation in Figure S6 in that the bias is primarily caused by the pressure variable $x_{21}$, albeit with a smaller magnitude. As before, the prior misspecification for the $CO_2$ profile near the surface (levels 18, 19 and 20) causes little harm due to the small bias multipliers of those variables (Figure 2(a)).

Overall, Figures S6–S8 show how challenging it is to understand, predict and explain the uncertainty quantification performance of the operational retrieval method. The method can

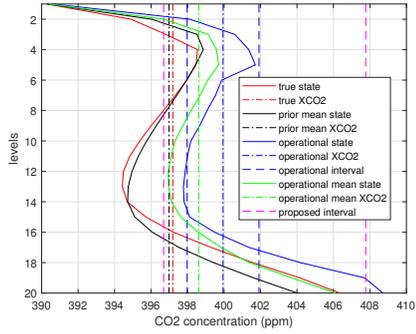
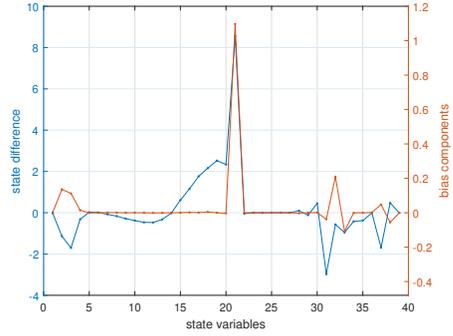

(a) Profile comparisons

(b) Bias components

**Figure S6:** Figure (a) illustrates the operational and proposed intervals for the state realization that has the smallest coverage in Figure 3(b). Figure (b) visualizes the corresponding state differences $x_i - \mu_{a,i}$ and bias components $m_i(x_i - \mu_{a,i})$, where $m_i$ is the $i$th bias multiplier.

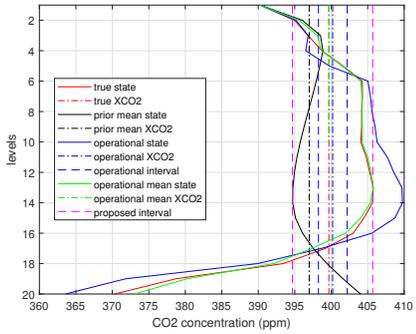
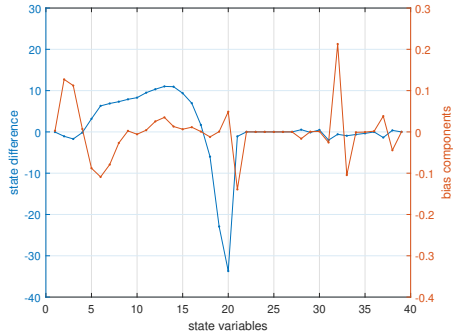

(a) Profile comparisons

(b) Bias components

**Figure S7:** Figure (a) illustrates the operational and proposed intervals for the state realization that has the largest coverage in Figure 3(b). Figure (b) visualizes the corresponding state differences $x_i - \mu_{a,i}$ and bias components $m_i(x_i - \mu_{a,i})$, where $m_i$ is the $i$th bias multiplier.

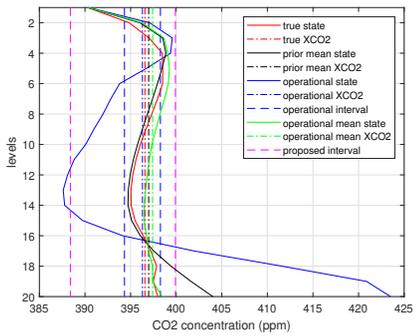
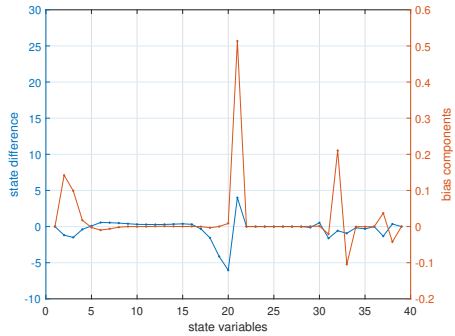

(a) Profile comparisons

(b) Bias components

**Figure S8:** Figure (a) illustrates the operational and proposed intervals for a state realization that has nominal coverage in Figure 3(b). Figure (b) visualizes the corresponding state differences $x_i - \mu_{a,i}$ and bias components $m_i(x_i - \mu_{a,i})$, where $m_i$ is the $i$th bias multiplier.

exhibit the very counterintuitive behavior where a relatively well-specified prior $CO_2$ profile (Figure S6) in fact has the worst coverage performance, while a badly misspecified prior $CO_2$ profile (Figure S7) has the highest coverage. The key to understanding the performance of the method, it turns out, is to understand the effect of the nuisance variables and how they interact with the misspecification of the $CO_2$ profile. Such analysis is obviously only possible when the true $x$ is known, which would make it very difficult to perform a similar study for real-life operational retrievals. The proposed method, on the other hand, is free from these complications and exhibits consistent 95% coverage for all $x$ realizations we have investigated.

## S4 Supplement to Section 5.3

Figure S9 shows the spatial coverage and bias patterns for a case where the state vector realizations are such that all 64 intervals across the region have coverage below the nominal value.

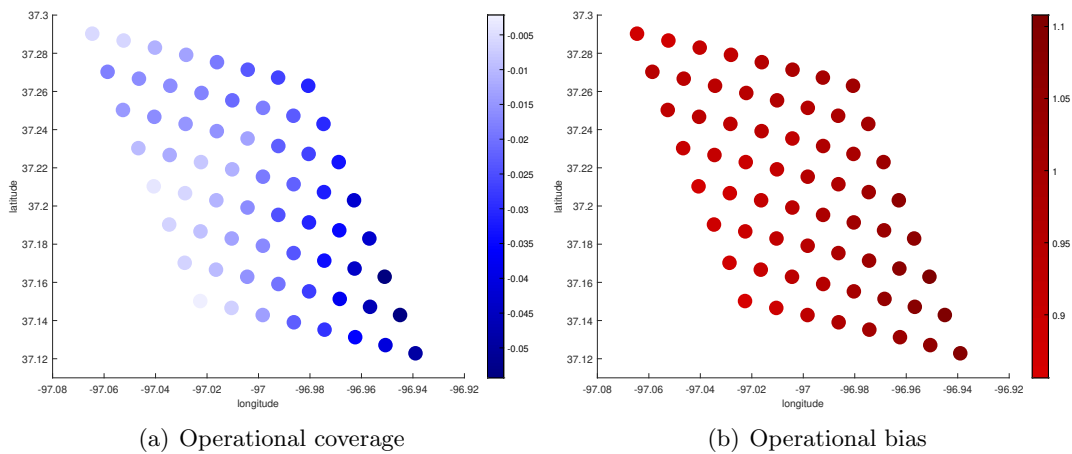

(a) Operational coverage  (b) Operational bias

**Figure S9:** Operational retrieval over a grid of $8 \times 8$ soundings for an instance with undercoverage for the entire grid. Figure (a) shows the spatial coverage pattern relative to the nominal 95% in units of probability (i.e., -0.03, for example, corresponds to coverage 0.92, instead of the nominal 0.95). Figure (b) shows the corresponding bias pattern in ppm.



\end{document}